\documentclass[%
 aip,
 amsmath,amssymb,
 reprint,%
]{revtex4-1}
\usepackage{amsmath}
\usepackage{amssymb}
\usepackage{color}
\usepackage{graphicx}
\usepackage{tabularx}
\usepackage{dcolumn} 
\usepackage{bm}      
\usepackage{array}   
\usepackage{amsfonts}
\usepackage{url}
\usepackage[bookmarks=false,colorlinks,citecolor=red]{hyperref}

\usepackage[utf8]{inputenc}
\usepackage[T1]{fontenc}
\usepackage{mathptmx}

\include{MyCommand}
\definecolor{purple}{rgb}{0.82, 0.23, 0.82}

\DeclareUnicodeCharacter{2212}{-}

\begin{document}

\newcommand{\cidag}[1]{\hat{c}_{#1}^{\dag}}
\newcommand{\ci}[1]{\hat{c}_{#1}}
\newcommand{\bidag}[1]{\hat{b}_{#1}^{\dag}}
\newcommand{\bi}[1]{\hat{b}_{#1}}
\newcommand{\aidag}[1]{\hat{a}_{#1}^{\dag}}
\newcommand{\ai}[1]{\hat{a}_{#1}}
\newcommand{\Ni}[1]{\hat{n}_{#1}}
\newcommand{\hatAphi}{\hat{A}(\vec{\phi})}
\newcommand{\hatAphiup}{\hat{A}_{\uparrow}(\vec{\phi})}
\newcommand{\hatAphidn}{\hat{A}_{\downarrow}(\vec{\phi})}
\newcommand{\Aphi}{A(\vec{\phi})}
\newcommand{\Aphiup}{A_{\uparrow}(\vec{\phi})}
\newcommand{\Aphidn}{A_{\downarrow}(\vec{\phi})}
\newcommand{\effspec}[1]{\tilde{\epsilon}_{#1}}
\newcommand{\overlap}[2]{\langle #1 | #2 \rangle}
\newcommand{\obs}[1]{\langle #1 \rangle}
\newcommand{\NUp}{N_{\uparrow}}
\newcommand{\NDn}{N_{\downarrow}}

\preprint{AIP/123-QED}

\title{Finite Temperature Auxiliary Field Quantum Monte Carlo in the Canonical Ensemble}

\author{Tong Shen}
 \affiliation{Department of Chemistry, Brown University, Providence, RI 02912
 }

\author{Yuan Liu}
 \affiliation{Center for Ultracold Atoms, Research Laboratory of Electronics, Massachusetts Institute of Technology, Cambridge, Massachusetts 02139
 }

\author{Yang Yu}
\affiliation{
Department of Physics, University of Michigan, Ann Arbor, MI 48109
}
\author{Brenda M. Rubenstein}
 \email{Author to whom correspondence should be addressed: brenda\_rubenstein@brown.edu.}
 \affiliation{Department of Chemistry, Brown University, Providence, RI 02912
 }

\date{\today}
\begin{abstract}
Finite temperature auxiliary field-based Quantum Monte Carlo methods, including Determinant Quantum Monte Carlo (DQMC) and Auxiliary Field Quantum Monte Carlo (AFQMC), have historically assumed pivotal roles in the investigation of the finite temperature phase diagrams of a wide variety of multidimensional lattice models and materials. Despite their utility, however, these techniques are typically formulated in the grand canonical ensemble, which makes them difficult to apply to condensates like superfluids and difficult to benchmark against alternative methods that are formulated in the canonical ensemble. Working in the grand canonical ensemble is furthermore accompanied by the increased overhead associated with having to determine the chemical potentials that produce desired fillings. Given this backdrop, in this work, we present a new recursive approach for performing AFQMC simulations in the canonical ensemble that does not require knowledge of chemical potentials. To derive this approach, we exploit the convenient fact that AFQMC solves the many-body problem by decoupling many-body propagators into integrals over one-body problems to which non-interacting theories can be applied. We benchmark the accuracy of our technique on illustrative Bose and Fermi Hubbard models and demonstrate that it can converge more quickly to the ground state than grand canonical AFQMC simulations. We believe that our novel use of HS-transformed operators to implement algorithms originally derived for non-interacting systems will motivate the development of a variety of other methods and anticipate that our technique will enable direct performance comparisons against other many-body approaches formulated in the canonical ensemble.
\end{abstract}


\maketitle

\section{Introduction}

For decades, chemical physicists have focused on developing a constellation of electronic structure methods, from mean field\cite{szabo1996modern,shavitt2009many} to perturbation\cite{szabo1996modern,shavitt2009many} to coupled cluster theories,\cite{shavitt2009many,Bartlett_RMP_2007} to describe the ground and low-lying excited states of molecules and solids. While it is undoubtedly true that many phenomena involve electrons that reside in their ground states, it is becoming increasingly apparent that there are a wealth of phenomena for which this assumption does not hold: atoms and molecules in the centers of large planets and stars can experience GPa of pressure and temperatures of over 10,000 K,\cite{Lorenzen_2014,McMahon_RMP_2012} and lasers can be used to heat and thereby steer chemical reactions along different mechanistic pathways to facilitate processes such as catalysis.\cite{Mukherjee_NanoLett,Zhou_NanoLett,Yong2020} Temperature is moreover one of the key parameters that can be used to tune the electronic properties of the materials that make up many of modern society's most important technologies\cite{Imada_RMP_1998,Orenstein468,Si2016} and is responsible for the pernicious loss of coherence in physical realizations of qubits.\cite{Ladd2010}

Reflecting this growing list of applications, a growing number of methods have recently been developed to study them. One of the most straightforward ways of determining the finite temperature properties of quantum systems is by diagonalizing the system's Hamiltonian to determine all of its many eigenvalues and weighting them to compute its partition function and related observables in a process termed exact diagonalization (ED).\cite{Jafari_ED} While this technique is numerically exact, as its name implies, its computational cost scales exponentially with system size. On the other end of the spectrum, finite temperature mean field theories, including finite temperature Hartree Fock\cite{MERMIN_1963,Sanyal1994} and Density Functional (DFT) theories,\cite{Karasiev_PRL_2014,Pribram_Jones_2014} trade accuracy for computational expediency by approximating the many-body problem as a one-body problem in which an electron is coupled to an average or ``mean'' field of the other electrons. While such mean field theories continue to improve and hold great promise, they still struggle to achieve the accuracy often needed to correctly capture many-body physics. 
Between these two poles lie a variety of techniques that are generalizations of their ground state counterparts. The past few years, for example, have seen a resurgence of interest in finite temperature coupled cluster techniques\cite{Harsha_JCP_2019,Harsha_JCTC_2019,White_Chan_2018, White_JCP_2020,Shushkov_JCP_2019} and  perturbation theories.\cite{Jha_2019,Jha_PRE_2020,White_arXiv,Hirata_2019,Hirata_Jha_2020,Santra_ChemPhys_2017} Closer to the mean field end of the spectrum, finite temperature embedding theories that partition systems into correlated regions embedded within uncorrelated baths\cite{zgid2017finite} such as Dynamical Mean Field Theory (DMFT)\cite{Georges_RMP,Kotliar_RMP_2006} and the finite temperature SEET and GF2 methods\cite{Welden_Sgid_JCP,Kananenka_JCTC,Kananenka_JCTC_2016_2} are distinctively capable of directly obtaining the full frequency-dependent spectra of the systems they model. Nevertheless, all of these methods struggle to balance computational cost with the need to account for the numerous electronic states that contribute to finite temperature expectation values. 

Finite temperature Quantum Monte Carlo (FT-QMC) methods are particularly advantageous in this regard because they have the exceptional ability to access numerous states without the exponential or high polynomial costs of other techniques by randomly sampling multidimensional state spaces for the most important states.\cite{Foulkes_RMP_2001,TOULOUSE_Umrigar,gubernatis2016quantum,Motta_WIRES} One of the most successful of these techniques is Determinant Quantum Monte Carlo (DQMC)\cite{White_PRB_1989,Hirsch_PRB,Bai_Chapter} and its more recent extension, FT Auxiliary Field Quantum Monte Carlo (FT-AFQMC).\cite{Zhang_PRL_1999,He_PRL_2019,Liu_JCTC_2018} DQMC has a long history of being employed to study a wide variety of lattice models. One of the key reasons why DQMC has been so widely adopted is because, for certain important classes of problems,\cite{Li_AnnRev,White_PRB_1989} it does not suffer from the infamous sign or phase problems\cite{Loh_PRB_1990} that limit the practical utility of many other stochastic methods. AFQMC methods grew out of DQMC methods and were developed with the aim of being able to accurately model systems with clear sign problems by leveraging its more sophisticated sampling techniques and phaseless approximation.\cite{Zhang_PRB_1997,Zhang_PRL_2003,Motta_WIRES,Zhang_PRL_1995} These methods have since shed light on such sign- and phaseful systems as the Hubbard model off of half-filling\cite{Loh_PRB_1990,Zheng1155,LeBlanc_PRX_2015} and many different \textit{ab initio} molecular\cite{Liu_JCTC_2018,Purwanto_JCP,Suewattana_PRB,Shee_JCTC} and solid state systems.\cite{Liu_JCTC_2020,Ma_PRL_2015,Purwanto_JCTC,Zhang_JCP} Furthermore, these QMC techniques are versatile -- they can be applied to virtually any second-quantized Hamiltonian -- and, by sampling the overcomplete space of non-orthogonal determinants, they are able to sample large portions of Fock space with a high accuracy, yet comparatively mild computational cost of $O(N^{3})$-$O(N^{4})$, where $N$ denotes the number of basis functions.\cite{Motta_WIRES}  

This said, one of the Janus-faced features of these methods is that they are formulated in the grand canonical ensemble, in which a system's internal energy and particle number are allowed to fluctuate according to its fixed temperature and chemical potential.\cite{mcquarrie2000statistical} In many systems, it is more natural to fix intensive properties rather than extensive quantities (imagine the inherent difficulty involved with fixing the number of electrons in a solid) and auxiliary field-based methods formulated in the grand canonical ensemble may therefore be viewed as being better suited for modeling these systems. Nevertheless, it can be challenging to ascribe a meaningful chemical potential to systems such as condensates, superfluids, and superconductors that, by definition,\cite{Annett:730995} can undergo large fluctuations in their particle numbers.\cite{Mullin_AJP} Indeed, for this very reason, past attempts at using grand canonical FT-AFQMC to model bosons and Bose-Fermi mixtures were largely unsuccessful.\cite{rubenstein2012finite} However, even for the majority of systems for which the chemical potential is well-defined, the process of determining the correct chemical potential to achieve a desired filling or average particle number can be unwieldy. In current auxiliary field techniques that work in the grand canonical ensemble, practitioners must scan through tens to hundreds of possible chemical potentials before recovering the desired one for every Hamiltonian they study, which can accumulate into a substantial cost. Moreover, sampling in the grand canonical ensemble is \textit{not} equivalent to sampling in the canonical ensemble, particularly at the particle numbers far from the thermodynamic limit used in many simulations. Sampling the grand canonical ensemble involves sampling a much larger Fock space of states, many of which minimally contribute to thermodynamic averages, than sampling the canonical ensemble. This makes comparing results from grand canonical ensemble approaches to results from canonical ensemble approaches challenging and can also introduce additional statistical noise which can make arriving at meaningful statistical averages more challenging than in the canonical ensemble. Because of the differences between these ensembles, properties measured in these ensembles may also converge to their limits in different manors, meaning that one ensemble may converge certain properties more rapidly than the other. 

Given this context, in this work, we derive a new recursive formulation for performing finite temperature AFQMC simulations in the canonical ensemble. Unlike past canonical ensemble formalisms that relied upon Fourier extracting canonical results from grand canonical simulations and thus also depended upon a costly tuning of chemical potentials,\cite{Ormand_PRC,Sedgewick_PRB_2003,Gilbreth_arXiv_2019,ROMBOUTS1998453,Wang_Toldin_PRE_2017} our technique does not require knowledge of the chemical potential. To accomplish this, our method exploits one of the key, yet often overlooked features of the AFQMC formalism: by decoupling two-body operators into an integral over one-body operators, the Hubbard-Stratonovich (HS) Transformation\cite{Hirsch_PRB_1983,Motta_WIRES,Buendia_PRB} used in AFQMC produces an ensemble of non-interacting systems to which theories developed for non-interacting systems can be applied. In particular, to derive our approach, we apply a recursive formalism for obtaining the partition functions of ideal gases to the one-body operators in AFQMC to ultimately yield a many-body recursive theory. In the following, we derive this formalism for systems of bosons, spinless fermions, and spinful fermions, and demonstrate its accuracy and practical advantages for several benchmark Hubbard models. Interestingly, we show that energies computed in the canonical ensemble converge more quickly to the ground state than energies computed in the grand canonical ensemble. Ultimately, we believe that this formalism will be most useful for studying condensates as well as systems of fermions approaching their ground states, and will motivate the integration of other useful non-interacting theories into the AFQMC formalism. 

We thus begin in Section \ref{methods} by deriving our formalism for bosons and fermions and contrasting it with the conventional grand canonical DQMC and AFQMC formalism. We then benchmark the accuracy and highlight interesting features of our method in Section \ref{sec:results}. Finally, in Section \ref{sec:conclusions}, we conclude with a discussion of the applications for which we expect our methodology to be of the greatest future use. 

\section{Methods \label{methods}} 

To contrast our canonical ensemble formalism with previous formalism, we begin by outlining 
the conventional grand canonical DQMC algorithm before describing our own formalism. 

\subsection{\label{sec:dqmc} Review of Determinant Quantum Monte Carlo (DQMC) in the Grand Canonical Ensemble} 
 
The central quantity in any grand canonical ensemble formalism is the grand partition function, $\mathcal{Z}$, because it is from this quantity that all other thermodynamic quantities can be obtained. In the grand canonical ensemble, the internal energy and particle number are allowed to fluctuate around average values that can be tuned by the temperature and chemical potential, respectively.\cite{mcquarrie2000statistical} The grand partition function may thus be expressed as the trace over the Boltzmann factor containing both the temperature and the chemical potential.

In DQMC\cite{White_PRB_1989,Bai_Chapter} and  AFQMC\cite{Zhang_PRL_2003,rubenstein2012finite} methods that work in the grand canonical ensemble, the grand partition function may be re-expressed into a form amenable to sampling by first discretizing it into $L$ imaginary time slices
\begin{equation}
    \mathcal{Z} = \text{Tr} \left(e^{-\beta (\hat{H} - \mu \hat{N})} \right) = \text{Tr} \left(\lim_{\Delta \tau \rightarrow 0} \prod_{l}^L e^{-\Delta \tau (\hat{H} - \mu \hat{N})}\right).
    \label{eqn:GrandPartitionFunction}
\end{equation}
Here, $\hat{H}$ denotes the many-body Hamiltonian, $\mu$ denotes the chemical potential, $\hat{N} = \sum_{i, \sigma} \cidag{i \sigma} \ci{i \sigma}$ denotes the particle number operator, and $\Delta \tau = \beta / L$. In order to facilitate its subsequent evaluation, the propagator, $e^{-\Delta \tau (\hat{H} - \mu \hat{N})}$, is factored into short imaginary time kinetic and potential propagators via a Suzuki-Trotter factorization\cite{Trotter,Suzuki_Prog1976} such that
\begin{equation}
    \mathcal{Z} \approx \text{Tr} \left( \prod_{l}^L [e^{-\Delta \tau \hat{K} / 2} e^{-\Delta \tau \hat{V} } e^{-\Delta \tau \hat{K} / 2}] \right).
    \label{eqn:S-T_Factorization}
\end{equation}
As discussed in Section \ref{benchmarks}, in this work, we focus on models whose Hamiltonians can be written as $\hat{H} = \hat{K} + \hat{V}$, where $\hat{K}$ denotes the collection of all one-body operators, including the chemical potential term, and $\hat{V}$ denotes that of all two-body operators. The exact grand partition function is recovered in the limit that ($\Delta \tau \rightarrow 0$). This factorization enables us to treat the one-body and two-body operators separately. While one-body propagators, $e^{-\Delta \tau \hat{K}}$, may be neatly expressed as matrices in a given basis,\cite{Hirsch_PRB} two-body propagators, $e^{-\Delta \tau \hat{V}}$, may not be as easily expressed. Fortunately, two-body propagators of the form $e^{- \Delta \tau \hat{V}}$ can be re-expressed in terms of one-body propagators (\textit{i.e.}, linearized).\cite{Hirsch_PRB_1983} In general, two-body operators such as $\hat{V}$ may be decomposed into combinations of squares of one-body operators\cite{Shiwei_Notes}
\begin{equation}
    \hat{V} = \sum_{\gamma} \lambda_{\gamma} \hat{v}_{\gamma}^2,
    \label{eqn:BilinearDecomposition}
\end{equation}
where the $\hat{v}_{\gamma}$ denote linear combinations of (and are therefore also linear) one-body operators and the $\lambda_{\gamma}$ weight each of those operators' contributions to the potential operator. 
Using such decompositions, these propagators may then be re-expressed as one-body propagators using the continuous Hubbard-Stratonovich (HS) Transformation\cite{Buendia_PRB}
\begin{equation}
    e^{-\Delta \tau \lambda_{\gamma} \hat{v}_{\gamma}^2} = \int_{-\infty}^{\infty} d\phi \frac{e^{- \phi^2 / 2}}{\sqrt{2 \pi}} e^{ \phi \sqrt{- \Delta \tau \lambda_{\gamma}} \hat{v}_{\gamma} }, 
    \label{eqn:HSTransformation}
\end{equation}
where $\phi$ is a so-called ``auxiliary field.'' As $e^{-\phi^{2}/2}/\sqrt{2\pi}$ is a Gaussian, it may be viewed as a probability and the overall transformation may be viewed as re-expressing a two-body propagator into an integral over effective, one-body mean field propagators parameterized by different auxiliary fields. Note that, while this continuous transformation is more general and can be used to decouple a wide variety of fermion and boson Hamiltonians,\cite{Motta_WIRES,Buendia_PRB,Shi_PRB_2013} it is often more efficient when treating Fermi Hubbard models to employ the discrete version of the HS Transformation.\cite{Hirsch_PRB_1983,Buendia_PRB}
Nevertheless, because it is more general, we will proceed to base our remaining derivations off of the continuous HS Transformation. 

Substituting Equation \eqref{eqn:HSTransformation} into Equation \eqref{eqn:GrandPartitionFunction}, the grand partition function may finally be written in a form that can be sampled
\begin{align}
    \mathcal{Z} &\approx \text{Tr} \Bigg( \prod_{l}^L e^{-\Delta \tau \hat{K} / 2} \int_{-\infty}^{\infty} d\Vec{\phi_l} p(\vec{\phi}_l) e^{ \sum_{\gamma} \phi_l^\gamma \sqrt{- \Delta \tau \lambda_{\gamma}} \hat{v}_{\gamma} } \nonumber \\
    & \hspace{3.5 cm} \times e^{-\Delta \tau \hat{K} / 2} \Bigg) \\
    &= \text{Tr} \left(\prod_{l}^L \int_{-\infty}^{\infty} d\Vec{\phi_l} p(\vec{\phi}_l) \hat{B}(\vec{\phi}_l) \right) \\
    &= \int_{-\infty}^{\infty} d\Vec{\phi_1} p(\vec{\phi}_1) \dots \int_{-\infty}^{\infty} d\Vec{\phi_L} p(\vec{\phi}_L) \text{Tr}\left(\prod_{l}^L \hat{B}(\vec{\phi}_l)\right),
    \label{eqn:Z_FinalForm}
\end{align}
as was our initial goal. In the above, $\vec{\phi}_l \equiv \{ \phi_l^\gamma \}$, the vector of all auxiliary fields, $p(\vec{\phi}_l)$ is the multivariate Gaussian distribution formed from the product of each field's Gaussian distribution, and
\begin{equation}
    \hat{B}(\vec{\phi}_l) = e^{-\Delta \tau \hat{K} / 2} e^{ \sum_{\gamma} \phi_l^\gamma \sqrt{- \Delta \tau \lambda_{\gamma}} \hat{v}_{\gamma} } e^{-\Delta \tau \hat{K} / 2}.
\end{equation}

One important advantage of working in the grand canonical ensemble is that, by taking the trace over fermions explicitly, the numerical complexities of evaluating $\text{Tr}(\prod_{l}^L \hat{B}(\vec{\phi}_l))$ are avoided and we obtain
\begin{multline}
    \mathcal{Z} \approx \int_{-\infty}^{\infty} d\Vec{\phi}_{1} p(\vec{\phi}_1) \dots \int_{-\infty}^{\infty} d\Vec{\phi_L} p(\vec{\phi}_L) \text{det} (I + \\ B_{\uparrow}(\vec{\phi}_L) \dots B_{\uparrow}(\vec{\phi}_1)) \text{det} (I + B_{\downarrow}(\vec{\phi}_L) \dots B_{\downarrow}(\vec{\phi}_1)),
    \label{eqn:ZDeterminantForm}
\end{multline}
where taking the trace over the product of the $\hat{B}(\vec{\phi}_l)$ leads to a product of determinants, one from each of the two spin sectors,\cite{Hirsch_PRB} $B_{\sigma}(\vec{\phi}_l)$ is the matrix representation of $\hat{B}_{\sigma}(\vec{\phi}_l)$ in the single-particle basis, and $I$ is the identity matrix. The grand partition function may thus be expressed as a multidimensional integral over auxiliary field space, which can be efficiently computed via Monte Carlo sampling.
It is based upon auxiliary field configurations sampled according to $\mathcal{Z}$ that observables such as energies and site/orbital occupancies may be determined. While this formalism has proven remarkably useful, it can be exceedingly costly to repeatedly determine the correct chemical potential at every temperature and system size one wants to model. 

\subsection{Obtaining Canonical Ensemble Properties from the Grand Canonical Formalism \label{sec:grandcantocan}} 
One can modify the grand canonical ensemble formalism described above to compute canonical ensemble quantities by replacing the Boltzmann factor containing the chemical potential term in Equation \eqref{eqn:GrandPartitionFunction} with $e^{-\beta \hat{H}}$ and taking the trace in Equation \eqref{eqn:Z_FinalForm} over all states in the \textit{canonical} ensemble instead of in the grand canonical ensemble. Nevertheless, evaluating the trace over canonical ensemble states
is far more complicated than evaluating the trace over grand canonical ensemble states as the particle number must be fixed. 

This said, several approaches have been advanced over the years to transform sampling of the grand canonical ensemble into sampling of the canonical ensemble without explicitly taking the trace over canonical ensemble states. One technique that takes advantage of grand canonical treatments while still fixing particle number is the particle projection method.\cite{Ormand_PRC,Sedgewick_PRB_2003,Gilbreth_arXiv_2019,ROMBOUTS1998453} In this technique, the grand partition function, as written in Equation \eqref{eqn:Z_FinalForm}, is multiplied by a constraint on the particle number in the form of a Kronecker delta, $\delta_{\hat{N}, N}$. It turns out that, by re-expressing the Kronecker delta as a Fourier transform and integrating, one can arrive at a version of Equation \eqref{eqn:ZDeterminantForm} in which the determinants are multiplied by a phase factor:\cite{Ormand_PRC}
\begin{equation}
    Z_N = \int d\vec{\phi} p(\vec{\phi}) \frac{1}{N_s} \sum_{m=1}^{N_s} e^{-i \psi_m N} \text{det}[1 + e^{i \psi_m}B(\vec{\phi})],
\end{equation}
where $\psi_m \equiv 2\pi m/N_s$ is the frequency of the Fourier transform.
As the key grand canonical equations remain intact, one can sample the grand canonical ensemble as before but with a chemical potential that constrains the particle number to fluctuate around a predetermined value. Nevertheless, the same need to scan through chemical potentials to target particle numbers as in the grand canonical formalism makes this approach  almost equally burdensome. A more direct canonical ensemble formalism is therefore crucial for boosting the computational efficiency, if not, feasibility, of many finite temperature auxiliary field simulations.

\subsection{\label{sec:bosons} Canonical Ensemble AFQMC Algorithm for Spinless Bosons and Fermions} 

Given this backdrop, we have approached this problem from a very different angle: motivated by the simple recursive relations that can be used to construct the canonical partition functions of ideal gases,\cite{borrmann1993recursion,borrmann1999calculation} we have derived an analogous set of recursive relations to construct canonical partitions functions in AFQMC. While such recursive relations are comparatively easy to construct for non-interacting systems because the energy spectra of such systems remain invariant under the addition/removal of particles, they are far harder to construct for interacting systems which possess many-body correlations. To surmount this fundamental problem, we have made use of the pivotal realization that the HS Transformation of Equation \eqref{eqn:HSTransformation} essentially transforms our interacting problems into integrals over non-interacting problems. Therefore, so long as we can perform an HS Transformation on a many-body Hamiltonian to decouple its interactions, we can then apply the same recursive relations that have been used for ideal gases to interacting systems. 

We begin by deriving our recursive formalism for the canonical partition function and then one- and two-body observables for bosons and spinless fermions. As we will show in the following derivations, our canonical ensemble formulation is the same for both types of species except for the different signs that arise in their final expressions due to their disparate counting statistics.

\subsubsection{Derivation of a Recursive Approach for Obtaining the Canonical Partition Function \label{partition}} 
Therefore, we denote $\hat{a}^{\dag}$ and $\hat{a}$ as particle creation and annihilation operators that can create/annihilate either fermions or bosons, respectively. The $N$-particle, canonical ensemble partition function may be expressed as
\begin{equation}
    Z_N = \text{Tr}_c (e^{-\beta \hat{H}}),
\end{equation}
which we can factor and transform just as we did the grand partition function to obtain
\begin{equation}
    \label{FirstZn}
    Z_N \approx \int_{-\infty}^{\infty} d\Vec{\phi_1} p(\vec{\phi}_1) \dots \int_{-\infty}^{\infty} d\Vec{\phi_L} p(\vec{\phi}_L) \text{Tr}_c (\prod_{l}^L \hat{B}(\vec{\phi}_l)). 
\end{equation}
In the above, we have added a subscript to the trace to differentiate it from that in Equation \eqref{eqn:Z_FinalForm}, as the trace can only be taken over all $N$-particle quantum states in the canonical ensemble.

Using basic commutation/anticommutation rules, it can be proven\cite{Hirsch_PRB} that the product of the $\hat{B}(\vec{\phi}_l)$ exactly corresponds to an exponential of some one-body operator parameterized by the auxiliary field vector $\vec{\phi} \equiv \{ \vec{\phi}_1, \dots, \vec{\phi}_L \}$, which we denote as $\hatAphi$
\begin{eqnarray}
     \label{Bphi} 
    \prod_{l}^L \hat{B}(\vec{\phi}_l) &=& \prod_{l}^L e^{-\Delta \tau \hat{K} / 2} e^{ \sum_{\gamma} \phi_l^\gamma \sqrt{- \Delta \tau \gamma_{\gamma}} \hat{v}_{\gamma} } e^{-\Delta \tau \hat{K} / 2} \nonumber \\ 
    &=& e^{- \beta \hatAphi}. 
\end{eqnarray}
We can subsequently define a set of new boson/fermion coordinates in which $\hatAphi$ is diagonal\cite{Hirsch_PRB}
\begin{equation}
    e^{- \beta \hatAphi} = e^{- \beta \sum_{i , j} \aidag{i} \Aphi_{ij} \ai{j} } = e^{- \beta \sum_{\gamma} \aidag{\gamma} \effspec{\gamma}(\vec{\phi}) \ai{\gamma} },
    \label{eqn:AExpansion}
\end{equation}
where we call the set of $\{ \effspec{\gamma}(\vec{\phi}) \}$ the effective single-particle spectrum, because of the following relation
\begin{equation}
    \Aphi = \sum_{i, j} | i \rangle \Aphi_{ij} \langle j | = \sum_{\gamma} | \gamma \rangle \effspec{\gamma} (\vec{\phi}) \langle \gamma |,
\end{equation}
and
\begin{equation}
    \aidag{\gamma} = \sum_{i} \overlap{i}{\gamma} \aidag{i}, \hspace{0.5 cm} \ai{\gamma} = \sum_{i} \overlap{\gamma}{i} \ai{i}.
    \label{eqn:BasisRotation}
\end{equation}
Since $e^{- \hatAphi}$ is an independent-particle propagator that only depends on the auxiliary field vector, $\vec{\phi}$, the effective single-particle spectrum, $\{ \effspec{\gamma}(\vec{\phi}) \}$, is independent of the particle number. For an $N$-particle, $N_s$-site system, taking the trace while constraining the particle number yields
\begin{align}
    \text{Tr}_c (e^{-\beta \hatAphi}) &= \text{Tr}_c (e^{- \beta \sum_{\gamma} \aidag{\gamma} \effspec{\gamma}(\vec{\phi}) \ai{\gamma} }) \label{eqn:CanonicalTrace1}\\
    &= \sum_{\Gamma} \langle \Gamma | e^{- \beta \sum_{\gamma} \aidag{\gamma} \effspec{\gamma}(\vec{\phi}) \ai{\gamma} } | \Gamma \rangle \label{eqn:CanonicalTrace2}\\
    &= \sum_{\Gamma} e^{- \beta \sum_{\gamma = 1}^{N_s} n_{\gamma} \effspec{\gamma}(\vec{\phi}) }. \label{eqn:CanonicalTrace3}
\end{align}
Here, $\Gamma$ is used to represent the set of $N$-particle states, and thus, $\sum_\Gamma \equiv \sum_{n_1 + \dots + n_{N_s} = N}$ and $n_\gamma$ denotes the number of particles in the $\gamma$th eigenstate.  For bosons, $n_\gamma = 0, 1, \dots, N$, whereas for fermions, $n_\gamma = 0, 1$. A more detailed derivation of these equations can be found in the Appendix. The key implication of Equation \eqref{eqn:CanonicalTrace3} is that, for a specified field $\vec{\phi}$, the single-particle spectrum can be decoupled from the particle number. Hence, the many-particle energy given such fields is simply the sum of all of the single-particle energies. 

This key fact enables us to move beyond previous projection-based approaches and calculate Equation \eqref{eqn:CanonicalTrace3} in a recursive fashion. To do so, we generalize the recursive approach to calculating canonical ensemble partition functions first developed for ideal gases.\cite{borrmann1993recursion, borrmann1999calculation} For an ideal gas of $N$ particles whose total energy can be written as the sum of single-particle energies $\{ \epsilon_i \}$, we can express $Z_N$ in terms of its subsystem partition functions, $Z_{N-k}$,
\begin{equation}
    \label{ZnOriginal}
    Z_N = \frac{1}{N} \sum_{k=1} (\pm 1)^{k+1} z_k Z_{N-k},
\end{equation}
with $Z_0 = 1$ and $z_k$, which can be identified as the partition function of the system at temperature $k\beta$, given by
\begin{equation} 
    z_k  = \sum_{i} e^{- k \beta \epsilon_{i}}. 
\end{equation} 
Because we can write our decoupled many-body energy in terms of its single-particle energies, just as in the ideal case, we can then denote $\text{Tr}_c (e^{-\beta \hatAphi})$ as $Z_N (\vec{\phi})$ and re-express it using the same recursive formalism 
\begin{equation}
    \label{Zn} 
    Z_N (\vec{\phi}) = \frac{1}{N} \sum_{k=1} (\pm 1)^{k+1} z_k (\vec{\phi}) Z_{N-k}(\vec{\phi}),
\end{equation}
with the same starting value $Z_0 (\vec{\phi}) = 1$ and
\begin{equation} 
    z_k (\vec{\phi}) = \sum_{\gamma} e^{- k \beta \effspec{\gamma}(\vec{\phi}) }.
\end{equation} 
In Equations \eqref{ZnOriginal} and \eqref{Zn}, the plus sign should be employed whenever Bose statistics are imposed, while the minus sign should be employed whenever Fermi statistics are imposed. Otherwise, our formalism is identical for systems of bosons and spinless fermions, since these systems involve only a single particle type and the total particle number can be constrained by a single value of $N$. Interestingly, this formalism bears a strong resemblance to a recursive formalism recently leveraged to incorporate particle statistics into Path Integral Molecular Dynamics simulations.\cite{Hirshberg_PNAS, Hirshberg_JCP} However, whereas that formalism was derived with real space path integrals in mind, ours was designed with an eye towards Fock space. 

Combining Equations \eqref{FirstZn}, \eqref{Bphi}, and \eqref{Zn}, we thus arrive at our final recursive expression for the many-body partition function 
\begin{eqnarray}
     Z_N &\approx& \int d\Vec{\phi} p(\vec{\phi}) Z_N (\vec{\phi}), \nonumber \\
     &=&  \int d\Vec{\phi} p(\vec{\phi}) \frac{1}{N} \sum_{k=1} (\pm 1)^{k+1} z_k (\vec{\phi}) Z_{N-k}(\vec{\phi}), 
     \label{eqn:ExactZN}
\end{eqnarray}
which demonstrates that the many-body partition function may be recovered by taking the integral over auxiliary field space of all recursively-constructed, field-dependent partition functions. As is customary, this integral may be evaluated through Monte Carlo sampling. 

\subsubsection{Derivation of a Recursive Approach for Obtaining Canonical Ensemble Observables \label{observables}} 

In the grand canonical ensemble, most observables of interest can be easily expressed in terms of elements of the one-body density matrix using thermal Wick's theorem.\cite{Fetter} Unfortunately, thermal Wick's theorem cannot be applied to products of field operators in the canonical ensemble.\cite{schonhammer2017deviations} In this Section, we therefore present an alternative approach to obtaining the expectation values of such products in the canonical ensemble based upon their occupation numbers, which is in the spirit of Sch\"{o}nhammer's derivations for non-interacting fermions.\cite{schonhammer2000thermodynamics}

In the canonical ensemble, the $(i, j)$-th element of the one-body density matrix may be expressed as
\begin{equation}
    \obs{ \aidag{i} \ai{j} } = \frac{ \text{Tr}_c (\aidag{i} \ai{j} e^{-\beta \hat{H}} ) }{ \text{Tr}_c (e^{-\beta \hat{H}}) }.
\end{equation}
Since the denominator can be explicitly evaluated via Equation \eqref{eqn:ExactZN}, we must determine an approach for evaluating the numerator, the unnormalized version of the one-body density matrix element, which we denote as $\tilde{D}_{ij}$. We can proceed to re-express the numerator as we did the partition function by performing an imaginary-time breakup and HS transformation of the propagators, which leads to 
\begin{eqnarray}
    \tilde{D}_{ij} &=& \text{Tr}_c (\aidag{i} \ai{j} e^{-\beta \hat{H}} ) \\
    &\approx& \int d\Vec{\phi} p(\vec{\phi}) \text{Tr}_c (\aidag{i} \ai{j} e^{-\beta \hatAphi }) \\
    &=& \int d\Vec{\phi} p(\vec{\phi}) \sum_{\Gamma} \langle \Gamma | \aidag{i} \ai{j} | \Gamma \rangle e^{- \beta \sum_{\gamma = 1}^M n_{\gamma} \effspec{\gamma}(\vec{\phi}) } \label{Dij_First}.
\end{eqnarray}
As is done in Equation \eqref{eqn:BasisRotation}, we can simplify this Equation by applying a change of basis to $\cidag{i} \ci{j}$ to yield
\begin{equation}
    \aidag{i} \ai{j} = \sum_{\lambda, \mu} \overlap{\lambda}{i} \overlap{j}{\mu} \aidag{\lambda} \ai{\mu} = \sum_{\lambda, \mu} U_{\lambda \mu}^{ij} \aidag{\lambda} \ai{\mu},
\end{equation}
where $U_{\lambda \mu}^{ij} \equiv \langle \lambda | i \rangle \langle j | \mu \rangle$ is an overlap matrix. This formula allows the integrand to be expressed as
\begin{eqnarray}
    \tilde{D}_{ij}(\vec{\phi}) &=& \sum_{\Gamma}  \sum_{\lambda, \mu} U_{\lambda \mu}^{ij} \langle \Gamma | \aidag{\lambda} \ai{\mu} | \Gamma \rangle e^{- \beta \sum_{\gamma = 1}^M n_{\gamma} \effspec{\gamma}(\vec{\phi}) }\\
    &=& \sum_{\Gamma} \sum_{\lambda} U_{\lambda \lambda}^{ij} \langle \Gamma | \aidag{\lambda} \ai{\lambda} | \Gamma \rangle e^{- \beta \sum_{\gamma = 1}^M n_{\gamma} \effspec{\gamma}(\vec{\phi}) } \\
    &=& \sum_{\lambda} U_{\lambda \lambda}^{ij} \obs{n_\lambda}_N Z_N(\vec{\phi}), \label{eqn:Dij_phi} \label{Dij_Second} 
\end{eqnarray}
where the second equality comes from the orthogonality of the eigenstates. We also define the mean occupation number, $n_{\lambda}$, of the $\lambda$th eigenstate with total particle number $N$ as 
\begin{equation}
    \obs{n_\lambda}_N = \frac{ \sum_{\Gamma} n_{\lambda} e^{- \beta \sum_{\gamma = 1}^M n_{\gamma} \effspec{\gamma}(\vec{\phi}) } } { \sum_{\Gamma} e^{- \beta \sum_{\gamma = 1}^M n_{\gamma} \effspec{\gamma}(\vec{\phi}) } },
    \label{one_particle} 
\end{equation}
which, as is illustrated in detail in the Appendix, can also be calculated in a recursive fashion
\begin{equation}
    \label{nlambda}
    \obs{\Ni{\lambda}}_N = \frac{Z_{N - 1}(\vec{\phi})}{Z_N(\vec{\phi})} e^{-\beta \epsilon_{\lambda}} \obs{1 \pm \Ni{\lambda} }_{N-1}.
\end{equation}
Here, the plus sign is again used to preserve Bose statistics, while the minus sign is used to preserve fermion statistics. As such, combining Equations \eqref{Dij_First}, \eqref{Dij_Second}, and \eqref{nlambda}, we arrive at the final expression for the one-body matrix elements
\begin{equation}
    \obs{ \aidag{i} \ai{j} } \approx \frac{ \int d\Vec{\phi} p(\vec{\phi}) [\sum_{\lambda} U_{\lambda \lambda}^{ij} \obs{n_\lambda}_N] Z_N(\vec{\phi}) }{\int d\Vec{\phi} p(\vec{\phi}) Z_N (\vec{\phi})}. \label{eqn:onebody_final}
\end{equation}

Following a similar scheme, the unnormalized $(i, j,  k, l)-$th element of the two-body density matrix can be expressed as
\begin{eqnarray}
    \tilde{D}_{ijkl} &=& \text{Tr}_c (\aidag{i} \ai{j} \aidag{k} \ai{l} e^{-\beta \hat{H}} ) \nonumber \\
    &\approx& \int d\Vec{\phi} p(\vec{\phi}) \tilde{D}_{ijkl}(\vec{\phi}),
\end{eqnarray}
where
\begin{eqnarray}
    \tilde{D}_{ijkl}(\vec{\phi}) &=& \big[ \sum_{\lambda \neq \nu} (U_{\lambda \lambda \nu \nu}^{ijkl} \pm U_{\lambda \nu \lambda \nu}^{ijkl} ) \obs{\Ni{\lambda} \Ni{\nu}}_N \\
    &\pm& \sum_{\lambda \neq \nu} U_{\lambda \nu \lambda \nu}^{ijkl} \obs{\Ni{\lambda}}_N + \sum_{\lambda} U_{\lambda \lambda \lambda \lambda}^{ijkl} \obs{\Ni{\lambda}^2}_N \big] Z_N(\vec{\phi}), \nonumber
\end{eqnarray}
\begin{equation}
    \obs{\Ni{\lambda}^2}_N = \frac{Z_{N-1}(\vec{\phi})}{Z_N(\vec{\phi})} e^{-\beta \epsilon_{\lambda}} \obs{(1 + \Ni{\lambda}) ^ 2 }_{N-1}, \label{eqn:nisquared}
\end{equation}
\begin{eqnarray}
    \obs{\Ni{\lambda} \Ni{\nu}}_N &=& \frac{Z_{N-1}(\vec{\phi})}{Z_N(\vec{\phi})} \big[\frac{e^{-\beta \epsilon_{\nu}}}{2} \obs{\Ni{\lambda} ( 1 \pm \Ni{\nu} )}_{N - 1} \nonumber \\
    &+& \frac{e^{-\beta \epsilon_{\lambda}}}{2} \obs{\Ni{\nu} ( 1 \pm \Ni{\lambda} )}_{N - 1} \big], \label{eqn:ninj}
\end{eqnarray}
and the change of basis is performed in the following way
\begin{eqnarray}
    \aidag{i} \ai{j} \aidag{k} \ai{l} &=& \sum_{\lambda, \mu, \nu, \xi} \overlap{\lambda}{i} \overlap{j}{\mu} \overlap{\nu}{k} \overlap{l}{\xi} \aidag{\lambda} \ai{\mu} \aidag{\nu} \ai{\xi} \nonumber \\
    &=& \sum_{\lambda, \mu, \nu, \xi} U_{\lambda \mu \nu \xi}^{ijkl} \aidag{\lambda} \ai{\mu} \aidag{\nu} \ai{\xi}.
\end{eqnarray}
Observe that the expression for $\obs{\Ni{\lambda}^2}_N$ only possesses a plus sign, which stems from the fact that two fermions cannot inhabit the same site. Again, the plus sign in the above equations corresponds to Bose statistics, while the minus sign coresponds to Fermi statistics.  The elements of higher order density matrices can be evaluated using the same basic scheme, which essentially requires manipulating the order of operators using commutation/anti-commuatation relations to create a sequence of occupation numbers that can be recursively calculated. With these one- and two-body matrix elements in hand, we can subsequently compute such important quantities as energies and correlation functions. 

\subsection{\label{sec:fermions} Canonical Ensemble AFQMC Algorithm for Spinful Fermions} 

Developing a canonical ensemble treatment for spinful fermion systems is naturally more difficult than developing a treatment for spinless fermion systems due to the additional constraint(s) that must be imposed on the partition function to constrain the spin in each spin sector. Moreover, more complicated anticommutation relations make generalizing our formalism to arbitrary spinful Hamiltonians, such as \textit{ab initio} Hamiltonians, far more challenging. Nevertheless, to simplify our discussion, here we focus on developing a formalism for treating the Fermi Hubbard Model given by Equation \eqref{eqn:Hamiltonian_Spinful} as a special, but important case given the model's long history in condensed matter physics. 

Since spin degrees of freedom may also be decoupled via an HS transformation for Hamiltonians of this type, Equation \eqref{eqn:Z_FinalForm} can be rewritten as
\begin{equation}
    Z_N \approx \int d\Vec{\phi} p(\vec{\phi}) \text{Tr}_c \big(\prod_{l}^L \hat{B}_{\uparrow}(\vec{\phi}_l) \prod_{l}^L \hat{B}_{\downarrow}(\vec{\phi}_l)\big),
\end{equation}
which means that spin up and spin down states are propagated separately, and we can accordingly define two one-body operators, $\hatAphiup$ and $\hatAphidn$, such that
\begin{equation}
    e^{-\beta \hatAphiup} = \prod_{l}^L \hat{B}_{\uparrow}(\vec{\phi}_l)
\end{equation}
and
\begin{equation} 
    e^{-\beta \hatAphidn} = \prod_{l}^L \hat{B}_{\downarrow}(\vec{\phi}_l).
\end{equation}
As a result, the partition function under a specified auxiliary field can be decomposed into the product of two subpartition functions with $N_\uparrow$ and $N_\downarrow$ particles, respectively
\begin{equation}
    Z_N(\vec{\phi}) = Z_{\NUp}(\vec{\phi}) \times Z_{\NDn}(\vec{\phi}),
\end{equation}
which implies that we can perform iterations over particle number for spin up and spin down partition functions separately. Thus, in line with our previous discussion, we end up with
\begin{multline}
    Z_N \approx \int d\Vec{\phi} p(\vec{\phi}) \big [\frac{1}{\NUp} \sum_{k=1} (- 1)^{k+1} z_k (\vec{\phi}) Z_{\NUp - k}(\vec{\phi}) \big] \\ \times \big[\frac{1}{\NDn} \sum_{k=1} (- 1)^{l+1} z_l (\vec{\phi}) Z_{\NDn - l}(\vec{\phi}) \big].
\end{multline}
Although expressions for the one- and two-body density matrix elements are more difficult to obtain for more general Hamiltonians, in the case of the Hubbard model, we can independently calculate matrix elements in the different spin sectors given a set of sampled auxiliary field vectors, $\vec{\phi}$. Thus, even though the field vectors that are ultimately sampled depend on both spin sectors, once these are obtained, the density matrix elements can be calculated separately. Hence, the unnormalized $(i, j)$-th elements of the spin up/down one-body matrices take the same form as Equation \eqref{eqn:Dij_phi}
\begin{equation}
    \tilde{D}_{ij}^{\sigma}(\vec{\phi}) = \sum_{\lambda} U_{\lambda \lambda, \sigma}^{ij} \obs{n_{\lambda \sigma}}_{N_\sigma} Z_{N_\uparrow}(\vec{\phi}) Z_{N_\downarrow}(\vec{\phi}),
\end{equation}
with the recursive relation being
\begin{equation}
    \obs{n_{\lambda \sigma}}_{N_\sigma} = \frac{Z_{N_\sigma - 1}(\vec{\phi})}{Z_{N_\sigma}(\vec{\phi})} e^{-\beta \epsilon_{\lambda \sigma}} \obs{1 - \Ni{\lambda \sigma} }_{N_\sigma - 1}. \label{eqn:ni_spinful}
\end{equation}
Additionally, the unnormalized $(i, j, k, l)$-th elements of the two-body density matrix may be obtained via
\begin{equation}
   \tilde{D}_{ijkl}(\vec{\phi}) = \sum_{\lambda, \mu} U_{\lambda \lambda, \uparrow}^{ij} U_{\nu \nu, \downarrow}^{kl} \obs{\Ni{\lambda_\uparrow}}_{N_\uparrow} \obs{\Ni{\nu_\downarrow}}_{N_\downarrow} Z_{N_\uparrow}(\vec{\phi}) Z_{N_\downarrow}(\vec{\phi}),
   \label{eqn:DMTwoBodySpin}
\end{equation}
where the change of basis is applied in the following way
\begin{align}
    \cidag{i \uparrow} \ci{j\uparrow} \cidag{k \downarrow} \ci{l \downarrow} &= \sum_{\lambda, \nu} \overlap{\lambda}{i} \overlap{j}{\lambda} \overlap{\nu}{k} \overlap{l}{\nu} \cidag{\lambda \uparrow} \ci{\lambda \uparrow} \cidag{\nu \downarrow} \ci{\nu \downarrow} \nonumber \\
    &= \sum_{\lambda, \nu} U_{\lambda \lambda, \uparrow}^{ij} U_{\nu \nu, \downarrow}^{kl} \cidag{\lambda \uparrow} \ci{\lambda \uparrow} \cidag{\nu \downarrow} \ci{\nu \downarrow}.
\end{align}
Note that, for the Hubbard model in which repulsions occur locally, only the $(i ,i ,i ,i)$-th elements of Equation \eqref{eqn:DMTwoBodySpin} have nonzero contributions.

\subsection{\label{sec:comp} Computational Details} 

Although our formulas for the field-dependent partition functions are exact, as written, these formulas are not necessarily computationally expedient to evaluate.    
First, the numerical effort to evaluate Equation \eqref{Zn} scales with the square of the particle number $N$. Moreover, $z_k(\vec{\phi})$ grows exponentially with the inverse temperature $\beta$ and the particle number $N$, which could cause severe numerical overflow issues at low temperatures.

To mitigate these issues, notice that in all mean occupation number recursions, namely Equations \eqref{nlambda}, \eqref{eqn:nisquared}, \eqref{eqn:ninj}, and \eqref{eqn:ni_spinful}, the partition functions appear as ratios. Hence, we may circumvent these computational challenges by calculating the ratios of these partition functions instead of the individual partition functions themselves. To achieve this, we may sum Equation \eqref{nlambda} over $\lambda$ and use the conservation of particle number to obtain
\begin{equation}
    N = \sum_{\lambda = 1}^{N_s} \obs{\Ni{\lambda}}_N = \frac{Z_{N - 1}(\vec{\phi})}{Z_N(\vec{\phi})} \sum_{\lambda = 1}^{N_s} e^{-\beta \epsilon_{\lambda}} \obs{1 \pm \Ni{\lambda} }_{N-1}.
\end{equation}
Some reordering yields
\begin{equation}
    \label{eqn:ZnRatio}
    \frac{Z_{N - 1}(\vec{\phi})}{Z_N(\vec{\phi})} = \frac{N}{\sum_{\lambda = 1}^{N_s} e^{-\beta \epsilon_{\lambda}} \obs{1 \pm \Ni{\lambda} }_{N-1}}.
\end{equation}
Observables may be obtained for each sampled field at a dramatically reduced cost based upon this equation as its cost only scales linearly with the particle number. 

In order to calculate observables that are expressed in terms of combinations of particle operators, we need to average appropriate combinations of density matrices through a Monte Carlo technique. As an illustration, we use an arbitrary one-body operator $\hat{T} = -\sum_{i,j} t_{ij} \aidag{i} \ai{j}$ and plug in Equation \eqref{eqn:onebody_final} to yield
\begin{eqnarray}
    \obs{\hat{T}} &=& -\sum_{i,j} t_{ij} \obs{\aidag{i} \ai{j}} \nonumber \\
    &\approx& \frac{ \int d\Vec{\phi} p(\vec{\phi}) [-\sum_{\lambda} t_{ij} U_{\lambda \lambda}^{ij} \obs{n_\lambda}_N] Z_N(\vec{\phi}) }{\int d\Vec{\phi} p(\vec{\phi}) Z_N (\vec{\phi})} \nonumber \\
    &\approx& \frac{1}{n} \sum_{\vec{\Phi}} [-\sum_{\lambda} t_{ij} U_{\lambda \lambda}^{ij} \obs{n_\lambda}_N],
\end{eqnarray}
with $n$ being the number of Monte Carlo samples and $\vec{\Phi}$ being sampled from a Boltzmann-like distribution
\begin{equation}
    P(\vec{\Phi}) = \frac{ p(\vec{\Phi}) Z_N(\vec{\Phi}) }{\int d\Vec{\phi} p(\vec{\phi}) Z_N (\vec{\phi})}.
\end{equation}
This can be accomplished by using standard sampling algorithms like the heat-bath or Metropolis algorithms.\cite{Landau_Binder}

In this paper, we elect to use the heat-bath algorithm to perform our integration over auxiliary fields because of its greater efficiency. If we define $R$ as the ratio of the new field-dependent partition function $Z_N(\vec{\Phi}')$ to the old partition function $Z_N(\vec{\Phi})$, the acceptance ratio of a random flip of discrete auxiliary fields or a random perturbation to continuous auxiliary fields may be written as 
\begin{equation}
    P = \frac{R}{1 + R}.
    \label{heat-bath}
\end{equation}

To benchmark this formalism against ED calculations and assess its performance relative to grand canonical ensemble algorithms, we implemented our own algorithms in Julia. Except where otherwise indicated, we set $\Delta \tau = 0.02$ in our calculations to yield accurate results comparable to those produced by ED, which we also implemented ourselves for both canonical and grand canonical ensemble calculations. In general, smaller $\Delta \tau$s lead to more accurate results. However, the effective single-particle spectrum, $\{ \effspec{\gamma}(\vec{\phi}) \}$, is $\Delta \tau$-dependent and becomes extremely large when $\Delta \tau$ is too small, exacerbating the numerical sign problem illustrated in Section \ref{sign_prob}. $\Delta \tau = 0.02$ thus is a balance point for us to investigate the models described in the following over physically meaningful parameter regimes without sign problems. All canonical ensemble AFQMC results are averaged over 50 walkers, which each sweep through the full lattice $2 \times 10^5$ times. The grand canonical ensemble results reported were obtained using a version of our previously-described finite temperature \textit{ab initio} AFQMC code modified to accommodate the Fermi Hubbard model.\cite{Liu_JCTC_2018} All of the codes that accompany this work can be found in our project Github repository.\cite{src}



\section{\label{sec:results} Results and Discussion}

\subsection{Benchmarks Against Exact Diagonalization Results for the Bose and Fermi Hubbard Models \label{benchmarks}}

To test the accuracy and analyze the computational performance of our canonical ensemble AFQMC (CE-AFQMC) algorithm, we benchmarked it against ED results for both the Bose and Fermi Hubbard models.  These models have long been used to illustrate and study the effects of strong correlation in materials, free of complicating material-specific details. As its name suggests, the Fermi Hubbard model (or, just the ``Hubbard'' model) exemplifies the two most basic behaviors of fermions, hopping and repulsion/attraction, on a lattice whose sites represent the locations of atoms within a material.\cite{Hubbard_Procs,Hubbard_Procs_2,Esslinger_AnnRev} The spinful version of this model is defined by the Hamiltonian
\begin{eqnarray}
    \hat{H} &=& \hat{K} + \hat{V} \nonumber \\
    &=& -t \sum_{\langle i, j \rangle, \sigma} (\cidag{i \sigma} \ci{j \sigma} + \text{h.c.}) + U \sum_{i} \hat{n}_{i \uparrow} \hat{n}_{i \downarrow},
    \label{eqn:Hamiltonian_Spinful}
\end{eqnarray}
where the first, one-body term describes the hopping of fermions from site $i$ to nearest-neighbor sites $j$ ($\langle ij \rangle$ denotes all pairs of nearest-neighbor sites $ij$), while the second term denotes the local, two-body fermion-fermion attraction/repulsion on the same site $i$. $t$ denotes the hopping parameter, which quantifies the relative magnitude of the hopping contribution to the Hamiltonian; $U$ is the attraction/repulsion constant which quantifies the relative magnitude of the two-body interaction to the Hamiltonian. If $U>0$, the fermion-fermion interaction is repulsive and the fermions are discouraged from occupying the same site; if $U<0$, the fermions are attracted to one another. $\uparrow$ and $\downarrow$ denote the up and down spins of the fermions. 

The Bose Hubbard model is a generalization of this model adapted to model bosons 
\begin{eqnarray}
    \hat{H} &=& \hat{K} + \hat{V} \nonumber \\
    &=& -t \sum_{\langle i, j \rangle} \bidag{i} \bi{j} + U \sum_{i} \Ni{i}^2,
    \label{eqn:Hamiltonian_Bose}
\end{eqnarray}
where $\bidag{i}$/$\bi{i}$ denote boson creation/annihilation operators at site $i$ and $\Ni{i} = \bidag{i}\bi{i}$ denotes the boson number operator. As before, $t$ and $U$ modulate the relative contributions of the boson hopping and interaction terms. In the past, this model has been employed to describe the physics of bosons in an optical lattices\cite{Greiner2002} and superfluids.\cite{Fisher_PRB_1989}

\begin{figure}
\centering{
\includegraphics[width=0.45\textwidth]{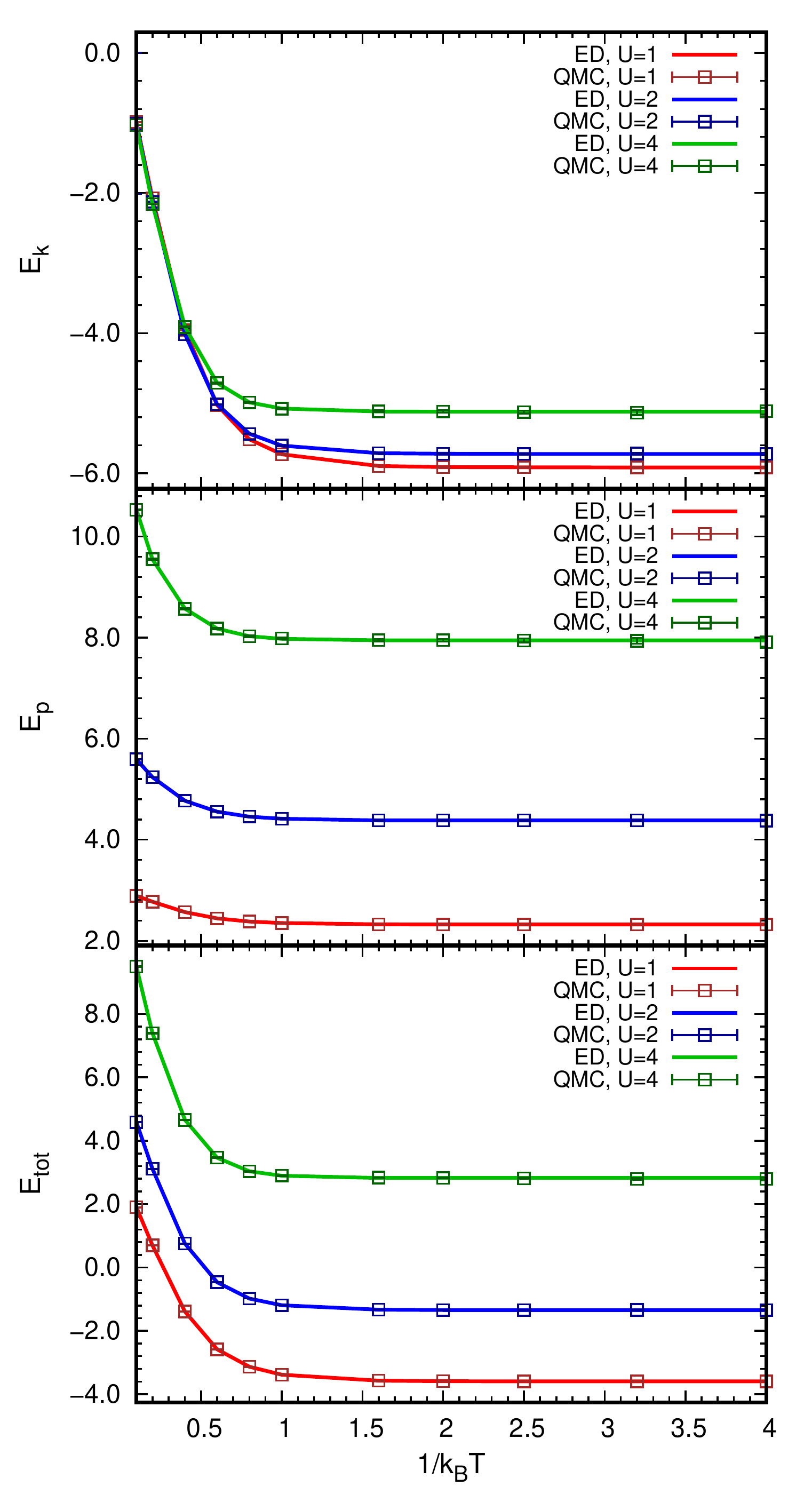}}
\caption{Total ($E_{\text{tot}}$), kinetic ($E_{\text{k}}$), and potential ($E_{\text{p}}$) energies of a three-site Bose-Hubbard model simulated for several values of $U$ at $t=1$ and $N = 3$. Energies are given in units of $t$.}
\label{BosonBenchmark}
\end{figure}
\begin{figure}
\centering{
\includegraphics[width=0.45\textwidth]{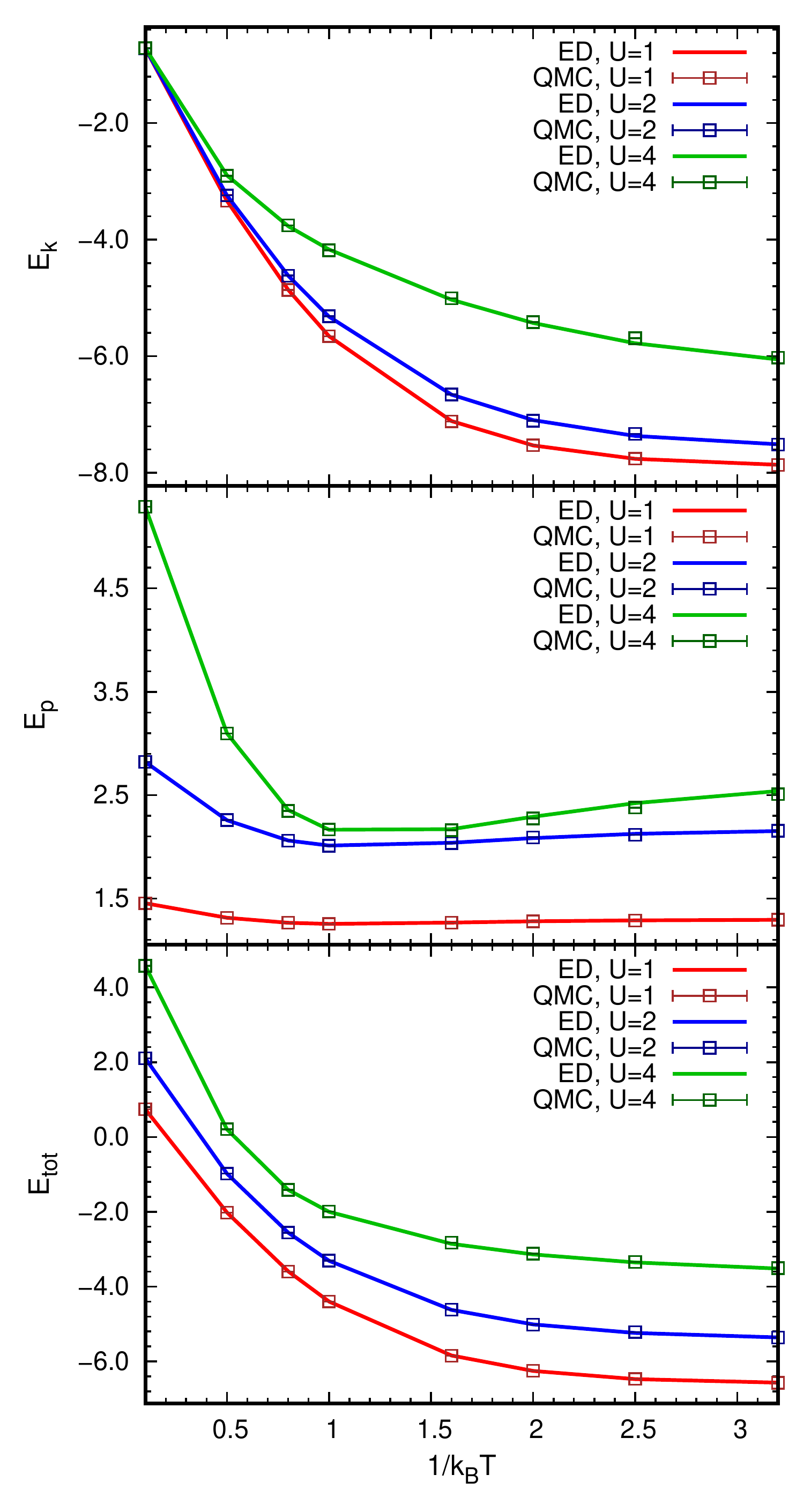}}
\caption{Total ($E_{\text{tot}}$), kinetic ($E_{\text{k}}$), and potential ($E_{\text{p}}$) energies of a six-site Hubbard model simulated for several values of $U$ at $t=1$ and $\NUp = \NDn = 3$. Energies are given in units of $t$.}
\label{FermionBenchmark}
\end{figure}

In Figure \ref{BosonBenchmark}, we compare our canonical ensemble AFQMC results against canonical ensemble results obtained using ED for a 1D, three-site Bose-Hubbard model containing three bosons, while in Figure \ref{FermionBenchmark}, we compare our CE-AFQMC and ED results for a 1D, six-site Hubbard model at half-filling. In order to obtain the total energies of these systems, we must compute their kinetic energies, $E_{k}$, which are one-body quantities, and potential energies, $E_{p}$, which are two-body quantities, which serve as direct tests of the formalism we presented for computing many-body density matrices in Sections \ref{sec:bosons} and \ref{sec:fermions}. For both systems, we make these comparisons for several different positive values of $U$ from high temperatures down to sufficiently low temperatures that their energies begin to plateau to their ground state values. It is apparent from these figures that our algorithm yields exact results, within statistical error bars, for both of these models across all of the $U$ values and temperatures surveyed. Direct numerical comparisons are presented in the tables in the Supplemental Materials. Interestingly, whereas the Bose-Hubbard results monotonically converge to their final plateaus with decreasing temperature, the Fermi Hubbard results manifest kinks at larger $U$ values in their potential energy curves between $0.75$ and $2$ $1/k_{B}T$, illustrating the competition that occurs between the model's kinetic and potential contributions.
These results verify that our formalism can achieve exact results using the heat bath algorithm described above over physically meaningful parameter regimes. 

\subsection{Emergence of the Sign Problem \label{sign_prob}} 

The results presented for the Fermi Hubbard model above naturally beg the question: how does the sign problem arise in this formalism? After all, the sign problem should arise in this QMC approach, just as it arises in all previous QMC approaches.\cite{Troyer_PRL_2005} We observe two origins of the sign problem in our formalism -- one ``physical'' and one ``numerical.'' The unavoidable ``physical'' origin of the sign problem in our approach stems from the fact that the expectation value of the propagator for the states of certain systems in Equation \eqref{eqn:CanonicalTrace3} can become negative, resulting in a mixture of positive and negative terms that can cancel one another out during sampling. This physical sign problem corresponds to that which emerges in grand canonical ensemble algorithms when the trace over grand canonical states yields products of determinants that can also be negative.\cite{Loh_PRB_1990} 

\begin{figure}
\centering{ 
\includegraphics[width=0.45\textwidth]{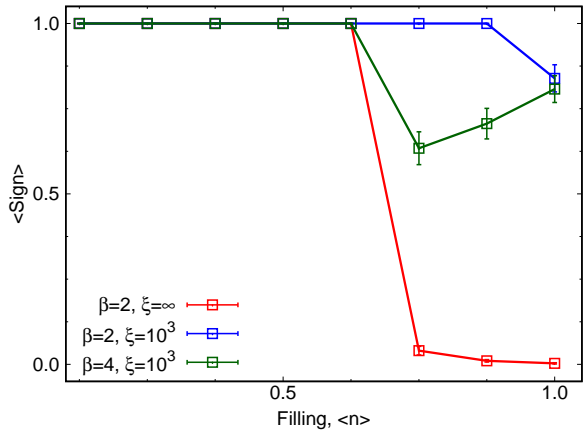}}
\caption{The average sign of $Z_N(\vec{\phi})$ as a function of filling for a 16-site, $U=4$ Hubbard model for various $\beta$s and $\xi$ cutoff values. $\xi = \infty$ means that no cut-off approximation is employed.}
\label{Sign}
\end{figure}

While we did not observe such a sign problem in the Fermi Hubbard simulations presented above, we did observe the second numerical form of the sign problem. This numerical sign problem arises from rounding errors that can accumulate during the calculation of Equation \eqref{nlambda} and has been reported in several previous papers describing recursive treatments of ideal Fermi gases.\cite{schonhammer2000thermodynamics, schonhammer2017deviations, magnus2019occupation} 
More specifically, when a large $e^{-\beta \epsilon_{\lambda}(\vec{\phi})}$ appears in the recursive calculation, its corresponding mean occupation number, $\obs{n_{\lambda}}_N$, tends to approach 1, yielding $\obs{1 - n_{\lambda}}_N$ values as small as $10^{-15}$. As illustrated in Table IV of the Supplemental Materials, when this occurs, the corresponding occupation number cannot be correctly represented by finite precision arithmetic and corrupts the recursion. As a result, there exists a critical particle number, $N_c$, at which $\obs{n_\lambda}_{N_c}$ begins to exceed 1 for the $\lambda$ that corresponds to the largest eigenvalue of $e^{-\beta \Aphi}$, which violates the Pauli Exclusion Principle and causes the ratio of partition functions in Equation \eqref{eqn:ZnRatio} to become negative in turn. This is illustrated for a 16-site Hubbard model in Figure \ref{Sign}, from which we can clearly see a rapid collapse of the average sign as the number of particles surpasses $N_c$ at $\beta=2$ and $\xi=\infty$ (meaning that no approximations were invoked). In contrast with the physical sign problem described above, if left untreated, this numerical sign problem renders the simulation either completely sign-problem-free or completely noisy as a function of filling. 

\begin{table}
    \centering
    \begin{tabular}{c|c|c|c}
    \hline \hline
     & ED & QMC, $\xi = 10^3$ & QMC, $\xi = 10^4$ \\
    \hline
    $N_s = 6, \obs{n} = 1$ & −3.6429 & -3.61(3) & -3.67(6) \\
    $N_s = 8, \obs{n} = 0.75$ & -6.6298 & -6.61(1) & -6.64(5) \\
    $N_s = 8, \obs{n} = 1$ & -4.5002 & -4.44(3) & -4.51(5) \\
    \hline \hline
    \end{tabular}
    \caption{Comparison of QMC calculations with different cutoff thresholds ($\xi$) against Exact Diagonalization (ED) results for the $U=4$ Hubbard model at $\beta = 5$. The $\lambda_l$th eigenstate is discarded in the recursive calculation if $\exp \left( -\beta(\epsilon_{N}(\vec{\phi}) -  \epsilon_{\lambda_l}(\vec{\phi}) ) \right) > \xi$ holds, where $N = \frac{1}{2}\obs{n}N_s$.}
    \label{tab:SpectrumTruncation}
\end{table}


To curb this problem, we have developed an approximation similar in spirit to the selection of active spaces in other electronic structure techniques that strikes a balance between alleviating the numerical sign problem and biasing the simulation. This approximation is motivated by the observation that the numerical instabilities our simulations incur stem from the fact that, at low temperatures, the $e^{-\beta \epsilon_{\lambda}(\vec{\phi})}$ become extremely large for small $\lambda$s and extremely small for large $\lambda$s. Thus, to prevent our results from being corrupted by round-off errors, the influence of these large and small $e^{-\beta \epsilon_{\lambda}(\vec{\phi})}$s on the other $e^{-\beta \epsilon_{\lambda}(\vec{\phi})}$s must be reduced. To accomplish this, notice that $\epsilon_N(\vec{\phi})$ is the energy level closest to the Fermi level and, at low temperatures, the occupancies of energy levels far below the Fermi level tend to be ``frozen'' around $1$. Based upon this observation, a coarse-grained scheme can be devised in which the occupancies of the energy levels $\epsilon_{\lambda_l}(\vec{\phi})$ (where $l$ denotes lower) are set to 1 ($\obs{n_{\lambda_l}}_{N}=1$) if $e^{-\beta(\epsilon_{N}(\vec{\phi}) -  \epsilon_{\lambda_l}(\vec{\phi}))} > \xi$, where $\xi$ is a cutoff value.  A similar approximation can then be made by setting $\obs{n_{\lambda_u}}_{N}=0$ for energy levels $\epsilon_{\lambda_u}(\vec{\phi})$ (where $u$ denotes upper) well above the Fermi level such that $e^{-\beta(\epsilon_{\lambda_u}(\vec{\phi}) -  \epsilon_{N}(\vec{\phi}))} > \xi$. The smaller the $\xi$, the smaller the numerical sign problem, but the larger the bias. Conversely, the larger the $\xi$, the larger the numerical sign problem, but the smaller the bias.

That said, we benchmarked the performance of this approximation against ED for the Hubbard model for different $\xi$ values at different temperatures and $U$s, as reported in Table \ref{tab:SpectrumTruncation}. As expected, a finer approximation, $\xi=10^4$, yields more accurate results, and with this cutoff value, less than 1\% of the samples are corrupted by the numerical sign problem. In contrast, a coarser approximation with $\xi=10^3$ completely eliminates the numerical sign problem for all parameter regimes we investigated, but its results are more biased. To better analyze the applicability of this scheme, we performed several simulations on the 16-site Hubbard model with $U=4$, and plotted the average sign of the partition function $Z_N(\vec{\phi})$ as function of filling, $\obs{n} = \frac{\NUp + \NDn}{N}$. We chose $\xi = 10^3$ to get rid of any effects from the numerical sign problem, in order to better investigate the emergence of physical sign problem in our algorithm. As illustrated in Figure \ref{Sign}, we find that when the cutoff approximation is applied, the overall sign problem is greatly improved at $\beta=2$, and the physical sign problem emerges at $\obs{n}=1$, which is a consequence of the fact that, as one iterates over the particle number, some extreme auxiliary field configurations are more likely to occur. For the $\beta=4$ curve, one finds that the physical sign problem emerges at $\obs{n}=0.75$ and decreases as the filling increases. 
These examples demonstrate that a reasonable choice of $\xi$ can mitigate the numerical sign problem without significantly biasing the algorithm's results, highlighting the algorithm's overall scalability.


\subsection{Convergence to the Ground State \label{convergence_ground}} 

One of the advantages that accompanies simulating in a different ensemble is that properties computed in that ensemble may converge to the ground state as a function of temperature in a different manner than in the original ensemble. Any differences in convergence to the ground state between the canonical and grand canonical ensembles would be particularly important to note in the context of AFQMC because, the higher the temperature at which properties approximate ground state properties, the more likely that those properties can be reliably modeled without a significant sign problem. To draw this comparison between ensembles, we therefore simulated the same six-site Fermi Hubbard model described in Section \ref{benchmarks} using both our CE and GCE \cite{rubenstein2012finite,Liu_JCTC_2018} algorithms from high to low temperatures at which the model's properties approach their ground state values. Note that, although CE and GCE properties may differ at finite temperatures, they should converge to the same ground state values. This is because, while states with particle numbers that differ from the average may be readily accessed in grand canonical simulations at finite temperatures, the step-like nature of the particle number vs. chemical potential curve that arises at low temperatures ensures that only states with a fixed $N$ are sampled in the grand canonical ensemble at temperatures approaching the ground state. This consequently ensures that the same ground state energy will be attained in both ensembles. 
\begin{figure}
\centering{ 
\includegraphics[width=0.45\textwidth]{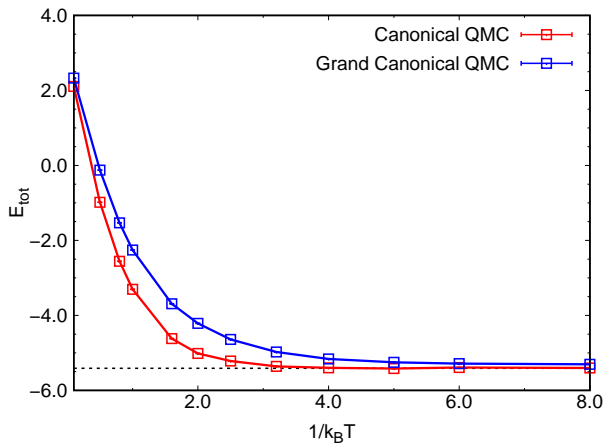}}
\caption{Six-site Fermi Hubbard model simulated within the canonical ensemble with $\NUp = \NDn = 3$ and the grand canonical ensemble with $\obs{N} = 6$ (half-filling) at $U=2$. $\Delta \tau$ is set to $0.05$ at $\beta = 8$ to avoid numerical sign problems. The horizontal dashed line at $E = -5.4095$ is the exact ground state energy.}
\label{EnsembleComparison}
\end{figure}
As illustrated in Figure \ref{EnsembleComparison} for $U=2$, we find that our CE-AFQMC algorithm converges significantly more rapidly than its GCE counterpart -- by about a factor of a few $1/k_{B}T$ units. Although this may seem small, a few $1/k_{B}T$ units can make the difference between being able to secure coveted insights into low-temperature physics or not. The intuitive reason why the canonical ensemble converges more rapidly is because, as alluded to above, it accesses fewer higher-energy states at any given temperature than the grand canonical ensemble. We expect this convergence difference to grow with increasing $U$, as evidenced in the Supplemental Materials. This finding suggests that our canonical algorithm may assume an essential role in future research focused on understanding how finite temperature algorithms and properties converge to the ground state -- especially given that ground state AFQMC algorithms are performed at fixed particle number. 

\subsection{Scaling Relative to the Grand Canonical AFQMC Algorithm \label{scaling}} 

Given our algorithm's encouraging ground state convergence properties, one may ask how it scales with sites and filling relative to the grand canonical algorithm. Because it relies upon the same imaginary time propagation of matrices as the DQMC and GCE-AFQMC algorithms before it,\cite{BSS_Algorithm,Zhang_PRL_1999} it requires $O(L N_s^3)$ operations to construct its full propagator. If we group all $L$ imaginary time steps into $m$ sets, the algorithm moreover requires $O(\frac{L^2}{m^2} N_s^3)$ operations for numerical stabilizations, as have previous algorithms.\cite{BSS_Algorithm} The distinguishing feature of our algorithm is its use of recursions. To initialize recursions, we must diagonalize the matrix $e^{-\beta \Aphi}$ and to account for $N$ particles, $O(N)$ recursions are required. These two operations thus involve an additional computational cost that scales as $O(N_s^3 + N)$. Therefore, unless $N \gg N_s$, the cost of performing recursions is negligible compared to the usual cost of propagating matrices. However, previous DQMC algorithms can rely on the Sherman-Morrison formula, which only requires $O(N_s^2)$ operations to update the determinant of a matrix with a rank-one perturbation, to quickly update their Green's functions.\cite{gubernatis2016quantum,White_PRB_1989} Because no analogous formula exists for updating the eigenvalues of a matrix with a rank-one perturbation, as is needed in our algorithm, the numerical cost of updating quantities after each Monte Carlo step in our approach scales as $O(N_s^3)$, making it the most expensive part of our algorithm.

\section{\label{sec:conclusions} Conclusions}

In this work, we have presented a new finite temperature AFQMC formalism for computing observables within the canonical ensemble. Unlike previous canonical ensemble algorithms, which relied upon projecting canonical ensemble properties out from grand canonical simulations, our formalism is based upon recursions among purely canonical ensemble quantities, eliminating the costly need to scan through chemical potentials to fix average particle numbers in most DMC and AFQMC simulations. To verify the accuracy of our approach while also demonstrating its broad applicability, we used our algorithm to calculate thermal properties of the Bose and Fermi Hubbard models, quintessential models of strong correlation, and benchmarked it against ED results. We moreover demonstrated that our canonical algorithm converges to ground states more rapidly with decreasing temperature than conventional grand canonical AFQMC algorithms, which illustrates its promise for studying how low temperature phenomena ultimately give rise to ground state phenomena. 

Although we presented a rather limited range of illustrations of our algorithm here, we foresee our algorithm having a wide variety of practical applications. First and foremost, because our canonical ensemble approach definitively fixes particle number, we believe that it will eliminate the `rogue eigenvalue problem' that curtails the direct application of grand canonical AFQMC algorithms to boson and fermion condensates whose average particle numbers become challenging to fix at low temperatures.\cite{rubenstein2012finite} This will enable the study of low temperature Bose and Bose-Fermi condensates\cite{rubenstein2012finite} as well as superconductors\cite{Gilbreth_PRA} with the high accuracy typical of AFQMC techniques. In addition, our algorithm will enable apples-to-apples, canonical ensemble comparisons between FT-AFQMC and other canonical ensemble finite temperature electronic structure techniques, such as Density Matrix QMC\cite{Petras_JCTC_2020} and finite temperature coupled cluster approaches,\cite{White_Chan_2018,White_JCP_2020,Harsha_JCP_2019,Harsha_JCTC_2019} for the first time. Some of these comparisons will necessitate generalizing our formalism to arbitrary \textit{ab initio} Hamiltonians, which is an ongoing effort. This will provide the community with the critical ability to compare and refine finite temperature methods, much as has been recently done for ground state electronic structure techniques.\cite{Williams_PRX,Motta_PRX_2017}

\section*{Supplementary Materials} 
Supplementary data, including additional benchmarks and an illustration of the numerical sign problem, may be found in the Supplementary Materials.

\section*{Data Availability} 
The data that support the findings of this study are also openly available in our Zenodo Repository at https://doi.org/10.5281/zenodo.3991899.\cite{Shen_Data}

\begin{acknowledgments}
The authors thank Richard Stratt, Hongxia Hao, James Shepherd, and members of the Center for Predictive Simulation of Functional Materials for stimulating conversations. T.S. received partial support from the Brown Open Graduate Education program, while Y.L. was graciously supported by the Brown Presidential Fellows program. B.R. acknowledges funding from the U.S. Department of Energy, Office of Science, Basic Energy Sciences, Materials Sciences and Engineering Division, as part of the Computational Materials Sciences Program and Center for Predictive Simulation of Functional Materials. Y.Y. contributed to this work while a part of the International Exchange Program for Excellent Students of University of Science and Technology of China. All calculations were conducted using computational resources and services at the Brown University Center for Computation and Visualization.
\end{acknowledgments}

\appendix
\section{Detailed Derivation of Partition Functions and Unnormalized Density Matrices}
\renewcommand{\theequation}{A\arabic{equation}}
In this Appendix, we provide detailed derivations of several recursive relations that are essential to obtaining the partition functions and density matrices for a specific auxiliary field. Some of these equations have been derived in other contexts before.\cite{hamann1990energy, rubenstein2012finite, schmidt1953einfache, schmidt1989simple}

First, we derive Equations \eqref{eqn:CanonicalTrace1}, \eqref{eqn:CanonicalTrace2}, and \eqref{eqn:CanonicalTrace3}, which undergird our expressions for the canonical partition function. Suppose we have an arbitrary $N \times N$ matrix $A$ expressed in the single-particle basis and form two Fock-space operators that operate on boson and fermion Fock states respectively
\begin{eqnarray}
    \hat{A}_{b} &=& \hat{b}^{\dag} A \hat{b}, \\
    \hat{A}_{f} &=& \hat{c}^{\dag} A \hat{c}.
\end{eqnarray}
We then consider the operations $\hat{A}_{b} \hat{b}^{\dag}$ and $\hat{A}_f \hat{c}^{\dag}$. Using the elementary commutation/anticommutation relations, it is straightforward to show that
\begin{eqnarray}
    \hat{A}_{b} \hat{b}_k^{\dag} &=& \sum_{ij} \hat{b}_i^{\dag} A_{ij} \hat{b}_j \hat{b}_k^{\dag} \nonumber \\
    &=& \sum_{i} \hat{b}_i^{\dag} A_{ik} + \hat{b}_k^{\dag} \sum_{ij} \hat{b}_i^{\dag} A_{ij} \hat{b}_j, \\
    \hat{A}_{f} \hat{c}_k^{\dag} &=& \sum_{ij} \hat{c}_i^{\dag} A_{ij} \hat{c}_j \hat{c}_k^{\dag} \nonumber \\
    &=& \sum_{i} \hat{c}_i^{\dag} A_{ik} - \sum_{ij} \hat{c}_i^{\dag} \hat{c}_k^{\dag} A_{ij} \hat{c}_j \nonumber \\
    &=& \sum_{i} \hat{c}_i^{\dag} A_{ik} + \hat{c}_k^{\dag} \sum_{ij} \hat{c}_i^{\dag} A_{ij} \hat{c}_j
\end{eqnarray}
which gives
\begin{eqnarray}
    \hat{A}_b \hat{b}^{\dag} &=& \hat{b}^{\dag} (A + I \hat{A}_b), \\
    \hat{A}_f \hat{c}^{\dag} &=& \hat{c}^{\dag} (A + I \hat{A}_f),
\end{eqnarray}
where $I$ is the $N \times N$ unit matrix. Note that the elements of the matrices $A$ and $I$ are $c$ numbers in Fock space and $\hat{A}$ is a scalar relative to the $N \times N$ matrices. As a result, $I \hat{A}$ and $A$ commute as matrices. Repeatedly applying these two equations yields
\begin{eqnarray}
    \hat{A}_b^n \hat{b}^{\dag} &=& \hat{b}^{\dag} (A + I \hat{A}_b)^n, \\
    \hat{A}_f^n \hat{c}^{\dag} &=& \hat{c}^{\dag} (A + I \hat{A}_f)^n
\end{eqnarray}
for any positive integer $n$. Hence,
\begin{eqnarray}
    e^{-\hat{A}_b} \hat{b}^{\dag} &=& \hat{b}^{\dag} e^{-(A + I \hat{A}_b)} = \hat{b}^{\dag} e^{-A} e^{-\hat{A}_b}, \label{eqn:BosonPropogation}\\
    e^{-\hat{A}_f} \hat{c}^{\dag} &=& \hat{c}^{\dag} e^{-(A + I \hat{A}_f)} = \hat{c}^{\dag} e^{-A} e^{-\hat{A}_f} \label{eqn:FermionPropogation}
\end{eqnarray}
can be shown inductively, where in the last step of both equations, the exponential is allowed to be broken up because the two terms in the exponent commute, as noted above. These results allow $e^{- \beta \hatAphi}$ in Equation \eqref{eqn:CanonicalTrace1} to ``walk through'' the string of creation operators that defines the $N$-particle state $| \Gamma \rangle$. More specifically, by expanding $| \Gamma \rangle$ in terms of particle creation operators, we have
\begin{eqnarray}
    e^{-\hatAphi} | \Gamma \rangle &=& e^{- \beta \sum_{\gamma} \aidag{\gamma} \effspec{\gamma}(\vec{\phi}) \ai{\gamma} } \prod_{\gamma} (\aidag{\gamma})^{n_{\gamma}} | 0 \rangle \nonumber \\
    &=& \left[\prod_{\gamma} (\aidag{\gamma} e^{- \beta \effspec{\gamma}(\vec{\phi})})^{n_{\gamma}}\right] e^{-\hatAphi} | 0 \rangle \nonumber \\
    &=& \left[\prod_{\gamma} (\aidag{\gamma} e^{- \beta \effspec{\gamma}(\vec{\phi})})^{n_{\gamma}} \right] | 0 \rangle,
\end{eqnarray}
where, in the last step, we have used the fact $e^{-\hatAphi} | 0 \rangle = | 0 \rangle$. Thus, sandwiching $e^{-\hatAphi}$ between $\langle \Gamma |$ and $| \Gamma \rangle$ yields
\begin{equation}
    \langle \Gamma | e^{-\hatAphi} | \Gamma \rangle = \prod_{\gamma}  e^{- \beta n_{\gamma} \effspec{\gamma}(\vec{\phi})} \label{eqn:TraceElements},
\end{equation}
which enables us to proceed from Equation \eqref{eqn:CanonicalTrace2}
to Equations \eqref{eqn:CanonicalTrace3}. Notice that Equations \eqref{eqn:BosonPropogation} and \eqref{eqn:FermionPropogation} have exactly the same form. As such, Equation \eqref{eqn:TraceElements} holds true for both boson and fermion propagators, as they propagate the corresponding states in the same manner.

Next, we show how to iteratively obtain the mean occupation numbers of the eigenstates, as well as the products of those occupancies, \textit{i.e.}, Equations \eqref{nlambda}, \eqref{eqn:nisquared}, and \eqref{eqn:ninj}. These are needed to evaluate a system's energy and correlation functions. Since bosons and fermions possess different particle statistics, we provide their derivations separately, starting with those for bosons.

To arrive at Equation \eqref{nlambda}, without loss of generality, we consider the $\lambda = 1$ case of Equation \eqref{one_particle}. By definition,
\begin{equation}
    \obs{\Ni{1}}_N = \frac{\sum_{\{ n_{\lambda} \}_1^{N}} n_{1} e^{-\beta \sum_{\gamma = 1}^{N_s} n_{\gamma} \epsilon_{\gamma}}}{Z_N},
    \label{recursion} 
\end{equation}
where we have introduced the shorthand $\{ n_{\gamma} \}_1^{N}$ to denote all states that conserve their particle number such that $\sum_{\gamma = 1}^{N_s} n_{\gamma} = N$ and have omitted all field dependence for clarity. Since $n_{1}$ can take values from $0$ up to $N$ for bosons, the above equation can be re-expressed as
\begin{eqnarray}
   \obs{\Ni{1}}_N Z_N &=& \sum_{n_{1} = 0}^N n_{1} e^{-\beta n_{1} \epsilon_{1}} \sum_{\{ n_{\gamma} \}_2^{N - n_1}} e^{-\beta \sum_{\gamma = 2}^{N_s} n_{\gamma} \epsilon_{\gamma}} \nonumber \label{nimean_1}\\
    &=& \sum_{n_{1} = 1}^N (1 + n_{1}) e^{-\beta (1 + n_{1}) \epsilon_{1}} \sum_{\{ n_{\gamma} \}_2^{N - n_1 - 1}} e^{-\beta \sum_{\gamma = 2}^{N_s} n_{\gamma} \epsilon_{\gamma}} \nonumber  \label{nimean_2}\\
    &=& e^{-\beta \epsilon_{1}} \sum_{\{ n_{\gamma} \}_1^{N - 1}} (1 + n_1) e^{-\beta \sum_{\gamma = 1}^{N_s} n_{\gamma} \epsilon_{\gamma}} \nonumber \label{nimean_3}\\
    &=& e^{-\beta \epsilon_{1}} \obs{1 + \hat{n}_1}_{N-1} Z_{N-1}, \label{nimean_4}
\end{eqnarray}
where in the first equality, we can perform the sum over $n_1$ first and leave the rest constrained by the particle number conservation constraint $\{ n_{\gamma} \}_2^{N - n_1} \equiv \sum_{\gamma = 2}^{N_s} n_{\gamma} = N - n_1$, while in the second equality, we have used the fact that the enumerations of $\{n_2, \dots, n_{N_s} \}$ are independent of $\{ n_1 = 0 \}$. This allows us to count $n_1$ from $n_1 = 1$ instead of $n_1 = 0$. Replacing $\hat{n}_1$ with $\hat{n}_1^2$ in the expectation values above gives
\begin{equation}
    \obs{\hat{n}_1^2}_N Z_N = e^{-\beta \epsilon_{1}} \obs{(1 + \hat{n}_1)^2}_{N-1} Z_{N-1}.
\end{equation}

The procedure leading to Equation \eqref{nimean_4} can easily be extended to calculate expectation values of products of different occupation numbers, \textit{i.e.}, $\obs{n_i n_j}_{N}$ with $i \neq j$. Replacing $\hat{n}_1$ with $\hat{n}_1 \hat{n}_2$ and observing that the enumerations of $\{n_3, \dots, n_{N_s} \}$ are independent from $\{ n_1 = 0, n_2 =0 \}$, one obtains
\begin{equation}
    \obs{\hat{n}_1 \hat{n}_2}_N = \frac{Z_{N-2}}{Z_N} e^{-\beta (\epsilon_{1} + \epsilon_{2})} \obs{(1 + \hat{n}_1) (1 + \hat{n}_2)}_{N-2}, \label{ninjmean_appendix}
\end{equation}
which is equivalent to Equation \eqref{eqn:ninj} as the indices in that equation are dummy variables and hence the equation can be rewritten into a symmetric form.

Since $n_{\gamma}$ can only be $0$ and $1$ for fermions, we arrive at the following two identities
\begin{eqnarray}
    \sum_{n_{\gamma} = 0,1} n_{\gamma} e^{-\beta n_{\gamma} \epsilon_{\gamma}} &=& e^{-\beta \epsilon_{\gamma}} \\
    \sum_{n_{\gamma} = 0,1} (1 - n_{\gamma}) e^{-\beta n_{\gamma} \epsilon_{\gamma}} &=& 1. 
\end{eqnarray}
Substituting these identities into the recursive relation for the mean occupation number, Equation \eqref{recursion}, yields
\begin{align}
    \obs{\Ni{1}}_N Z_N &= e^{-\beta \epsilon_{1}} \sum_{\{ n_{\gamma} \}_2^{N - 1}}^{n_{\gamma} = 0, 1} e^{-\beta \sum_{\gamma = 2}^{N_s} n_{\gamma} \epsilon_{\gamma}} \nonumber \\
    &= e^{-\beta \epsilon_{1}} \sum_{\{ n_{\gamma} \}_2^{N - 1}}^{n_{\gamma} = 0, 1} (1 - n_1) e^{-\beta \sum_{\gamma = 1}^{N_s} n_{\gamma} \epsilon_{\gamma}} \nonumber \\
    &= e^{-\beta \epsilon_{1}} \obs{1 - \hat{n}_1}_{N - 1} Z_{N - 1}
\end{align}
Proceeding along similar lines yields the fermion version of Equation \eqref{ninjmean_appendix}
\begin{equation}
    \obs{\hat{n}_1 \hat{n}_2}_N = \frac{Z_{N-2}}{Z_N} e^{-\beta (\epsilon_{1} + \epsilon_{2})} \obs{(1 - \hat{n}_1) (1 - \hat{n}_2)}_{N-2}.
\end{equation}
These are precisely the equations referenced in the text. 

\bibliography{ref}{}
\end{document}


\newcommand{\cidag}[1]{\hat{c}_{#1}^{\dag}}
\newcommand{\ci}[1]{\hat{c}_{#1}}
\newcommand{\bidag}[1]{\hat{b}_{#1}^{\dag}}
\newcommand{\bi}[1]{\hat{b}_{#1}}
\newcommand{\aidag}[1]{\hat{a}_{#1}^{\dag}}
\newcommand{\ai}[1]{\hat{a}_{#1}}
\newcommand{\Ni}[1]{\hat{n}_{#1}}
\newcommand{\hatAphi}{\hat{A}(\vec{\phi})}
\newcommand{\hatAphiup}{\hat{A}_{\uparrow}(\vec{\phi})}
\newcommand{\hatAphidn}{\hat{A}_{\downarrow}(\vec{\phi})}
\newcommand{\Aphi}{A(\vec{\phi})}
\newcommand{\Aphiup}{A_{\uparrow}(\vec{\phi})}
\newcommand{\Aphidn}{A_{\downarrow}(\vec{\phi})}
\newcommand{\effspec}[1]{\tilde{\epsilon}_{#1}}
\newcommand{\overlap}[2]{\langle #1 | #2 \rangle}
\newcommand{\obs}[1]{\langle #1 \rangle}
\newcommand{\NUp}{N_{\uparrow}}
\newcommand{\NDn}{N_{\downarrow}}

\title{Supplemental Information for ``Finite Temperature Auxiliary Field Quantum Monte Carlo in the Canonical Ensemble''}

\author{Tong Shen}
\affiliation{Department of Chemistry, Brown University, Providence, RI 02912}

\author{Yuan Liu}
 \affiliation{Center for Ultracold Atoms, Research Laboratory of Electronics, Massachusetts Institute of Technology, Cambridge, Massachusetts 02139
 }

\author{Yang Yu}
\affiliation{
Department of Physics, University of Michigan, Ann Arbor, MI 48109
}
\author{Brenda M. Rubenstein}
\affiliation{Department of Chemistry, Brown University, Providence, RI 02912}

\date{\today}

\maketitle

\section{Discrete Hubbard-Stratonovich Transformation in the Algorithm}
\label{HSTransformation}
In the following, we more clearly state how the Hubbard-Stratonovich (HS) Transformation is employed in the algorithm. We have used the continuous HS transformation for the Bose Hubbard model and the discrete HS transformation for the Fermi Hubbard model. While working with the continuous HS transformation has been described in the main text, we provide clear information on how the algorithm works with the discrete HS Transformation. In the discrete case, the integrals over auxiliary fields turn into sums over the spin field, $s_{i, l}$,
\begin{equation}
    Z_N(\vec{s}) \approx \sum_{s_{i, l} = \pm 1} \text{Tr}_c \big(\prod_{l}^L \hat{B}_{\uparrow}(\vec{s}_l) \prod_{l}^L \hat{B}_{\downarrow}(\vec{s}_l)\big),
\end{equation}
with
\begin{equation}
    \hat{B}_{\sigma}(\vec{s}_l) = \hat{B}_{\sigma}(s_{N_s, l}) \hat{B}_{\sigma}(s_{N_s - 1, l}) \cdots \hat{B}_{\sigma}(s_{1, l}).
\end{equation}
Hence, the fields, $\vec{S}$, are sampled from the discretized Boltzmann-like distribution
\begin{equation}
    P(\vec{S}) = \frac{Z_N(\vec{S})}{ \sum_{\{\vec{s}\} = \{ -1, 1\}^{N_s L} } Z_N(\vec{s}) },
\end{equation}
with $N_s L$ being the size of the imaginary-time-space mesh.

\section{Supplemental Data and Figures}
In the following, all of the data used to generate the figures in the main text are provided. 

\begin{table}[h!]
    \centering
    \begin{tabular}{c|c|c|c|c|c|c}
    \hline \hline
         \multirow{2}{1em}{$\beta$} & \multicolumn{2}{c|}{$U=4$} & \multicolumn{2}{c|}{$U=2$} & \multicolumn{2}{c}{$U=1$}\\
         \cline{2-7}
             & ED & AFQMC & ED & AFQMC & ED & AFQMC \\
         \hline
         0.1 & 9.4981 & 9.496(0) & 4.5847 & 4.584(5) & 1.8994 & 1.899(3)\\
         0.2 & 7.3900 & 7.393(8) & 3.1102 & 3.111(1) & 0.6951 & 0.695(1)\\
         0.4 & 4.6606 & 4.65(6) & 0.7550 & 0.754(2) & -1.3900 & -1.389(2)\\
         0.6 & 3.4650 & 3.46(6) & -0.4635 & -0.462(5) & -2.5829 & -2.583(2)\\
         0.8 & 3.0386 & 3.03(5) & -0.9801 & -0.980(4) & -3.1345 & -3.134(3)\\
         1.0 & 2.8967 & 2.90(3) & -1.1908 & -1.190(4)& -3.3812 & -3.381(4)\\
         1.6 & 2.8280 & 2.83(5) & -1.3308 & -1.330(7) & -3.5722 & -3.571(5)\\
         2.0 & 2.8254 & 2.82(8) & -1.3407 & -1.342(3) & -3.5903 & -3.590(8)\\
         2.5 & 2.8250 & 2.81(6) & -1.3427 & -1.34(4) & -3.5950 & -3.594(5)\\
         3.2 & 2.8250 & 2.79(9) & -1.3430 & -1.33(7) & -3.5958 & -3.596(1)\\
         4.0 & 2.8250 & 2.79(6) & -1.3431 & -1.34(4) & -3.5959 & -3.595(6)\\
    \hline \hline
    \end{tabular}
    \caption{The internal energy of the 3-site Bose-Hubbard model at various inverse temperatures with $N=3$, using ED (exact) and canonical ensemble AFQMC. All energies are reported in units of $t$.}
    \label{tab:BoseHubbard}
\end{table}

\begin{table}[h!]
    \centering
    \begin{tabular}{c|c|c|c|c|c|c}
    \hline \hline
         \multirow{2}{1em}{$\beta$} & \multicolumn{2}{c|}{$U=4$} & \multicolumn{2}{c|}{$U=2$} & \multicolumn{2}{c}{$U=1$}\\
         \cline{2-7}
             & ED & AFQMC & ED & AFQMC & ED & AFQMC \\
         \hline
         0.1 & 4.5753  &  4.574(8) & 2.10511 & 2.1051(4) & 0.73786 & 0.7378(5)\\
         0.5 & 0.2020  &  0.207(6) & -0.9832 & -0.9817(7) & -2.02080 & -2.0206(4)\\
         0.8 & -1.4120  &  -1.416(2) & -2.5567 & -2.557(3) & -3.5948 & -3.598(1)\\
         1.0 & -2.0021  &  -1.993(1) & -3.3083 & -3.303(6) & -4.4002 & -4.400(0)\\
         1.6 & -2.8581  &  -2.826(4) & -4.6206 & -4.618(4) & -5.8434 & -5.851(8)\\
         2.0 & -3.1378  &  -3.120(4) & -5.0121 & -5.014(4) & -6.2522 & -6.249(2)\\
         2.5 & -3.3533  &  -3.341(3) & -5.2410 & -5.216(0) & -6.4710 & -6.469(4)\\
         3.2 & -3.5160  &  -3.510(2) & -5.3571 & -5.359(6) & -6.5680 & -6.572(2)\\
    \hline \hline
    \end{tabular}
    \caption{The internal energy of the 6-site Fermi Hubbard model at various inverse temperatures with $\NUp = \NDn =3$, using ED (exact) and canonical ensemble AFQMC. All energies are reported in units of $t$.}
    \label{tab:FermiHubbard}
\end{table}

\begin{table}[h!]
    \centering
    \begin{tabular}{c|c|c}
    \hline \hline
    $\beta$ & CE-AFQMC & GCE-AFQMC \\
    \hline
    0.1 & 2.1051(5) & 2.327(2) \\
    0.5 & -0.9817(8) & -0.125(4) \\
    0.8 & -2.557(4) & -1.531(1) \\
    1.0 & -3.303(6) & -2.25(6) \\
    1.6 & -4.618(4) & -3.69(0) \\
    2.0 & -5.014(4) & -4.21(3) \\
    2.5 & -5.216(0) & -4.64(0) \\
    3.2 & -5.359(6) & -4.97(6) \\
    4 & -5.398(1) & -5.160(9) \\
    5 & -5.416(7) & -5.25(1) \\
    6 & -5.390(6) & -5.285(4) \\
    8 & -5.402(7) & -5.30(4) \\
    \hline \hline
    \end{tabular}
    \caption{The internal energy of the 6-site Fermi Hubbard model at various inverse temperatures, using the canonical ensemble AFQMC (CE-AFQMC) with $\NUp = \NDn = 3$ and the grand canonical ensemble AFQMC (GCE-AFQMC) with $\obs{N} = 6$. All energies are reported in units of $t$.}
    \label{tab:EnsembleComparison}
\end{table}

\begin{figure}[h!]
\centering{ 
\includegraphics[width=0.8\textwidth]{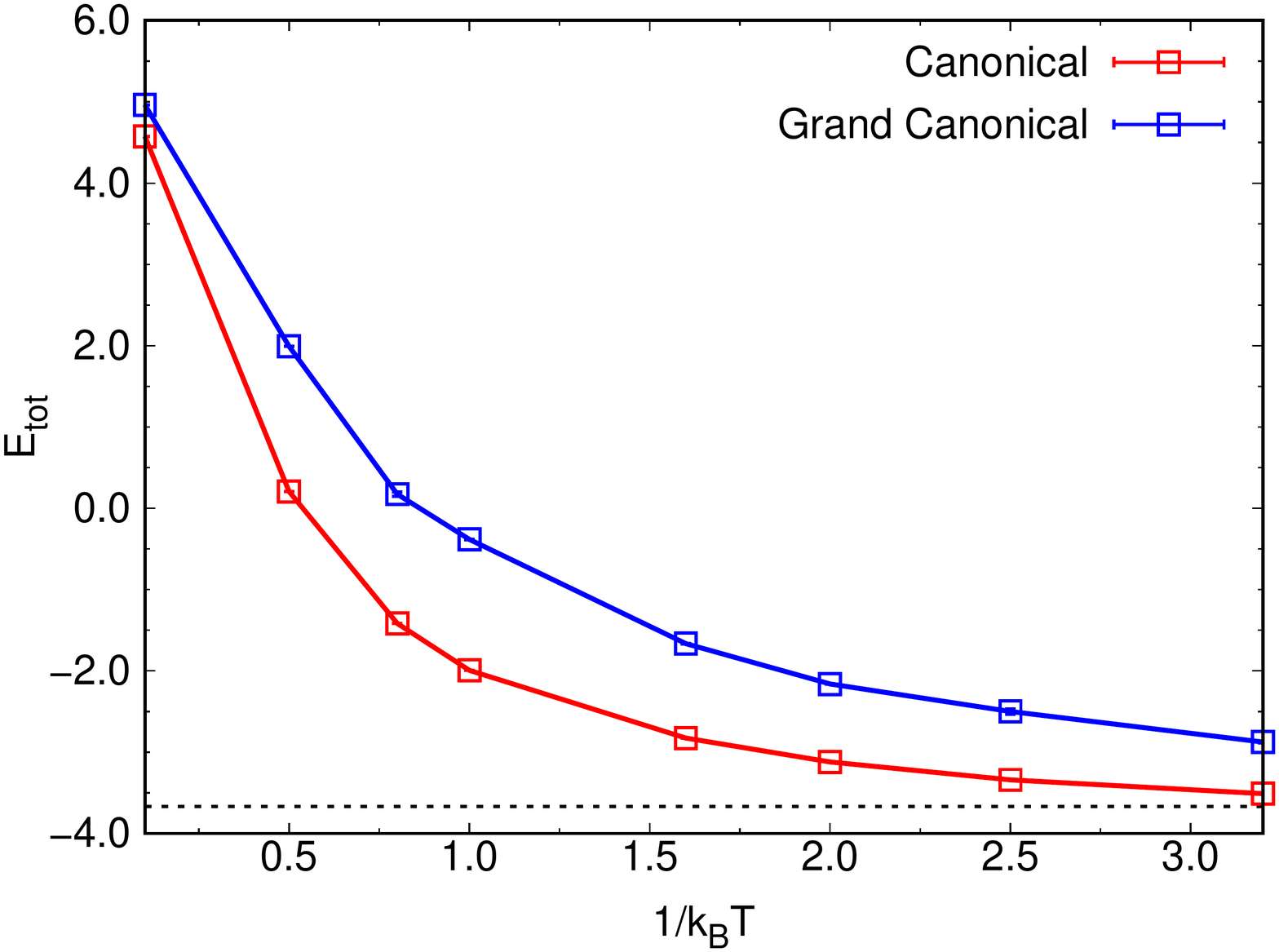}}
\caption{Six-site Fermi Hubbard model simulated within the canonical ensemble with $\NUp = \NDn = 3$ and the grand canonical ensemble with $\obs{N} = 6$ (half-filling) at $U=4$. The horizontal dashed line at $E = -3.6687$ is the exact ground state energy.}
\label{EnsembleComparison}
\end{figure}

\Edits{
\begin{table}
    \centering
    \begin{tabular}{c|c|c|c|c|c|c|c}
    \hline \hline
    $e^{-\beta \epsilon_{\lambda}(\vec{\phi})}$ & $\obs{n_\lambda}_4$ & $\obs{\tilde{n}_\lambda}_4$ & $\obs{n_\lambda}_6$ & $\obs{\tilde{n}_\lambda}_6$ & $\obs{n_\lambda}_7$& $\obs{\tilde{n}_\lambda}_7$ & $\obs{\tilde{n}_\lambda}_8$\\
    \hline
    7.9580E-11  &  1.6083E-11   & 0.0   &     8.6604E-10&   0.0     &   1.3524E-10  & 0.0   &0.0\\
    1.0294E-9   &  2.0805E-10   & 0.0   &     1.1203E-8 &   0.0     &   1.7494E-9   & 0.0   &0.0\\
    6.1206E-9   &  1.2370E-9    & 0.0   &     6.6609E-8 &   0.0     &   1.0401E-8   & 0.0   &0.0\\
    1.8378E-8   &  3.7142E-9    & 0.0   &     2.0000E-7 &   0.0     &    3.1232E-8  & 0.0   &0.0\\
    2.3903E-7   &  4.8310E-8    & 4.2103E-8 & 2.6014E-6 &   2.5942E-6  & 4.0622E-7  & 1.1723E-5 &1.6533E-4\\
    1.4734E-6   &  2.9778E-7    & 2.5952E-7 & 1.6034E-5 &   1.5990E-5  & 2.5039E-6  & 7.2262E-5 &0.001019\\
    8.3845E-6   &  1.6945E-6    & 1.4768E-6 & 9.1246E-5 &   9.0993E-5  & 1.4247E-5  & 4.1118E-4 &0.005797\\
    4.3413E-4   &  8.7739E-5    & 7.6467E-5 & 4.7230E-3 &   4.7100E-3  & 7.3429E-4  & 0.02119   &0.2939\\
    1.0922E-3   &  2.2073E-4    & 1.9237E-4 & 0.01187   &   0.01184 & 0.001834      & 0.05293   &0.7154\\
    0.02717     &  0.005487     & 0.004786  & 0.2896    &   0.2890  & 0.3280        & 0.9474    &0.9885\\
    0.06884     &  0.01388      & 0.01212   & 0.7103    &   0.7100  & 0.3388        & 0.9791    &0.9954\\
    2.0457      &  0.3881       & 0.3603    & 0.9902    &   0.9900  & 0.3386        & 0.9993    &0.9998\\
    3.5341      &  0.6389       & 0.6225    & 0.9944    &   0.9942  & 0.3333        & 0.9995    &0.9999\\
    31.14609    &  0.9581       & 1.0       & 0.9995    &   1.0     & 0.2313        & 1.0       &1.0\\
    302.1299    &  0.9957       & 1.0       & 1.0001    &   1.0     & -0.09100      & 1.0       &1.0\\
    3669.01383  &  0.9996       & 1.0       & 0.9989    &   1.0     & 6.9314        & 1.0       &1.0\\
    \hline \hline
    \end{tabular}
    \Edits{
    \caption{The illustration of the numerical sign problem for a 16-site, $U=4$ Hubbard model with filling $\obs{n}=1.0$. All results are based off of a single set of sampled auxiliary fields, $\vec{\phi}$. $e^{-\beta \epsilon_{\lambda}(\vec{\phi})}$ represents the set of statistical weights for each eigenstate and $\obs{n_\lambda}_i$ stands for the corresponding mean occupation number of each state, where $i$ denotes the total particle number. $\obs{n_\lambda}_i$ without the tilde is the occupation number \textit{without} the application of the spectral cutoff approximation, while $\obs{\tilde{n}_\lambda}_i$ denotes the same quantity, but with the approximation applied. The cutoff parameter is chosen to be $\xi = 10^4$, and since the eighth site is closest to the Fermi level at this filling, we have $\exp \left( -\beta(\epsilon_{8}(\vec{\phi}) -  \epsilon_{3}(\vec{\phi}) ) \right) > \xi$ and $\exp \left( -\beta(\epsilon_{13}(\vec{\phi}) -  \epsilon_{8}(\vec{\phi}) ) \right) > \xi$, meaning that the highest four energy levels and lowest three energy levels are discarded when iterating to obtain $\obs{\tilde{n}_\lambda}_i$. $\obs{\tilde{n}_\lambda}_i$ differs with $\obs{n_\lambda}_i$ at $i = 4$, but the discrepancies drop quickly with increasing iterations, as can be seen by the time $i=6$. Notice that the second lowest energy level (meaning second highest statistical weight) of $\obs{n_\lambda}_6$ is slightly greater than 1 due to the round-off error, which in turn corrupts the seventh iteration, $\obs{n_\lambda}_7$. However, by truncating the high and low energy levels, $\obs{\tilde{n}_\lambda}_i$ gives good results at higher $i$s. Moreover, since we are only using $\obs{\tilde{n}_\lambda}_8$ to calculate observables, errors at smaller $i$s are not significant.}}
    \label{tab:SpectrumTruncation}
\end{table}
}

\bibstyle{apsrev4-1}
\bibliography{ref}{}